\documentclass[12pt]{iopart}

\usepackage{fancyhdr}

\usepackage{iopams}
\usepackage{graphics,wrapfig,times}
\usepackage{graphicx}

\usepackage{amsfonts}
\usepackage{mathrsfs}
\usepackage[T1]{fontenc}

\begin{document}


\title{Cosmological dynamics with propagating Lorentz connection modes of spin zero}
\author{Hsin Chen$^a$, Fei-Hung Ho$^a$, James M. Nester$^{a,b,c}$,\\
Chih-Hung Wang$^a$, Hwei-Jang Yo$^d$}
\address{$^a$Department of Physics, National Central University,\\ 
No. 300, Jhongda Rd., Jhongli 320, Taiwan\\
$^b$Institute of Astronomy, National Central University,\\ 
No. 300, Jhongda Rd., Jhongli 320, Taiwan\\
$^c$Center of Mathematics and Theoretical Physics, 
National Central University,\\
No. 300, Jhongda Rd., Jhongli 320, Taiwan\\
$^d$Department of Physics, National Cheng-Kung University,\\
No. 1, University Rd., Tainan 701, Taiwan\\
E-mail: hchen@ntnu.edu.tw, 93242010@cc.ncu.edu.tw, nester@phy.ncu.edu.tw,\\
chwang@phy.ncu.edu.tw, hjyo@phy.ncku.edu.tw}


\begin{abstract}
The Poincar\'e gauge theory of gravity has a Lorentz
connection with both torsion and curvature.  For this theory  two
good propagating connection modes, carrying spin-$0^+$ and
spin-$0^-$, have been found. The possible effects of the spin-$0^+$
mode in cosmology were investigated in a previous work by our group;
there it was found that the $0^+$ mode could account for the
presently accelerating universe. Here, we extend the analysis to
also include the spin-$0^-$ mode. The resulting cosmological model
has three degrees of freedom. We present both the Lagrangian and
Hamiltonian form of the dynamic equations for this model, find the
late-time normal modes, and present some numerical evolution cases.
In the late time asymptotic regime the two dynamic modes decouple,
and the acceleration of the Universe oscillates due to the
spin-$0^+$ mode. 
\end{abstract}

\maketitle

\section{Introduction}






This work reports on an extension of a certain cosmological model, based on
the Poincar\'e gauge theory of gravity (PGT), which was first announced in
\cite{YN07} and then presented in considerable detail in
\cite{SNY08}.  In the latter work it was shown that the dynamic Riemann-Cartan
geometry (with curvature and torsion) could contribute an
oscillating aspect to the Universe expansion which could account for the
present day observed acceleration.
Since then two new works have appeared analyzing the
dynamics of this model and addressing its fit to the cosmological observations
\cite{LSX09,LSX09b}.  These works have already covered many
features of the original model in considerable detail.
Here we wish to first review the results of the application of certain
theoretical principles to the PGT.  That will naturally lead us to a more
appropriate description and our extension of the original model.

One of the outstanding successes of theoretical physics in the
latter part of the last century which led to a much deepened
understanding was the recognition that all the known fundamental
physical interactions, the strong, weak, and electromagnetic---{\it
not excepting gravity}---can be well described in terms of a single
unifying principle: that of local gauge theory. Although there are
other possible gauge approaches, for gravity it seems highly
appropriate to regard it a gauge theory for the local symmetry group
of Minkowski space time: the Poincar\'e group \cite{HHKN,Nes84}.
Such a consideration led to the development of the  Poincar{\'e}
Gauge Theory of gravity (PGT)
\cite{Hehl80,HS80,MieE87,HHMN95,GFHF96,Blag02}. The PGT has {\em a
priori} independent local rotation and translation gauge vector
potentials: the Lorentz (i.e., metric compatible) connection and the
orthonormal co-frame; their associated field strengths are the {\em
curvature} and {\em torsion}. The space-time then has generically a
Riemann-Cartan geometry. Because of its gauge structure and
geometric properties the PGT has been regarded as an attractive
alternative to general relativity.

The theory includes as exceptional cases Einstein's general
relativity (GR) with {\em vanishing} torsion, the Einstein-Cartan
theory with {\em non-dynamic} torsion algebraically coupled to the
intrinsic spin of the source, as well as the teleparallel
theories---wherein curvature vanishes but torsion does not.
The generic PGT has, in addition to the metric familiar from GR, a
connection with some independent dynamics, manifested in both the
torsion tensor and additional non-vanishing post-Riemannian
curvature components.

%
Investigations (especially \cite{HS80,SN80}) of the linearized
theory have identified six possible dynamic connection modes,
carrying certain spins and parity: $2^\pm,1^\pm,0^\pm$. It is not
possible for all of the modes to have good dynamics. The possible
combinations of well behaved (carrying positive energy at speed $\le
c$) propagating modes in the linear PGT theory were identified. The
Hamiltonian analysis revealed the related constraints \cite{BMNI83}.
Then detailed investigations of the Hamiltonian and propagation
\cite{HNZ96,CNY98,yo-nester-99,yo-nester-02} concluded that effects
due to nonlinearities in the constraints could be expected to render
all of these cases physically unacceptable except for the two
``scalar modes'', carrying spin-$0^+$ and spin-$0^-$.

One mode (referred to as the ``pseudoscalar'' because of its $0^-$
spin content) is reflected in the axial vector torsion. Axial
torsion is naturally driven by the intrinsic spin of fundamental
fermions; in turn it naturally interacts with such sources. Thus for
this mode one has some observational constraints
\cite{CSFG94,BSSI99}. Note that except in the early universe one
does not expect large spin densities. Consequently it is generally
thought that axial torsion must be small and have small effects at
the present time. The other good mode, $0^+$, the so-called
``scalar'' mode, is reflected in the vector torsion. There is no
known fundamental source which directly excites this mode.
Conversely this part of the connection does not interact in any
direct obvious fashion with any familiar type of matter
\cite{ShaI02}. Hence we do not have much in the way of constraints
as to its magnitude.  We could imagine it as having significant
magnitude and yet not being dramatically noticed---except indirectly
through the non-linear equations.

Thus the theoretical PGT analysis led to just two dynamic Lorentz
connection modes.   An obvious place where one might see some
physical evidence for these modes is in cosmological models. The
cosmological homogeneous and isotropic assumptions greatly restrict
the possible types of non-vanishing fields. Curiously, for the
connection and torsion there are only two possibilities, which
reflect precisely the two spin-0 connection modes. The scalar $0^+$
which gives rise to a special vector torsion which has only a time
component, and the pseudoscalar, $0^-$ mode, which gives rise to an
axial torsion which is the dual of a vector with only a time
component. Hence the homogeneous and isotropic cosmologies are {\em
naturally} very suitable for the exploration of the physics of the
dynamic PGT ``scalar modes''.

Thus cosmological models offer a situation where a dynamic Lorentz
connection may lead to observable effects.  Here we will not focus
on the early universe, where one could surely expect large effects
(although their signature would have to be disentangled from other
large effects), and instead inquire whether one could see any
effects of the PGT dynamic connection in the present day universe.
In particular we will here consider accounting for the outstanding
present day mystery: the accelerated universe, in terms of an
alternate gravity theory with an additional natural dynamic
geometric quantity: a Lorentz connection \cite{YN07,SNY08}.

The observed accelerating expansion of the Universe suggested the
existence of a kind of dark energy with a negative pressure. The
idea of a dark energy is one of the greatest challenges for our
current understanding of fundamental physics
\cite{PPRB03,PadT03,CoST06}. Among a number of possibilities to
describe this dark energy component, the simplest may well be by
means of a cosmological constant $\Lambda$.
%
Another popular
idea is the quintessence field --- some unusual type of minimally coupled
scalar field --- which has
received much attention over the
past few years and a considerable effort has been made in
understanding the role of quintessence fields on the dynamics of the
Universe (see, e.g., \cite{CaDS98,CarS98,BKBR05}).

An alternative is to consider some other gravity theory. Which
brings us to our specific topic: the possibility of explaining the
accelerating universe using a well tested alternative gravity theory,
one well motivated by both geometrical and physical gauge theory
principles. We explore the possibility that the dynamic PGT Lorentz
connection modes can drive the acceleration of the universe. As
noted above, there are two spin-0 modes which could have dynamical
behavior. In \cite{SNY08} it was shown  that the the spin $0^+$ mode
can make the expansion rate oscillate,
 naturally having an accelerating expansion in some periods and a
decelerating expansion at other times. For suitable choices of
parameters and initial data the model can account for the supernova
observations.  Here we show that including the $0^-$ mode allows for
an improved matching.

Over the years there have been many studies of PGT cosmology,
especially by Minkevich and coworkers (see, e.g.,
\cite{Min80,Min83,MN95,MG06,MGK07}).  Using various models they
found that it was possible for the PGT to avoid singularities,
account for inflation, and produce the acceleration of the universe
(as discussed later in Section~3, their mechanism is different from
that of our dynamic $0^+$ mode). A comprehensive early survey of the
PGT cosmological models was presented quite some time ago by Goenner
and M{\"u}ller-Hoissen \cite{GMH}. Although that work only solved in
detail a few particular cases, it developed the equations for all
the PGT cases---including those for the particular model we consider
here. However that work was done prior to the discovery of the
accelerating universe, and torsion was thus imagined as playing a
big role only at high densities in the early universe. More recently
investigators have begun to consider various models with torsion as
a possible cause of the accelerating universe (see,
e.g.,~\cite{MGK07,Boe03,CaCT03,WW09}).

We have taken another step in the exploration of the possible
evolution of the Universe with dynamic Lorentz connection spin-0
modes of the PGT. The main motivation is two-fold: (1) to have a
better understanding of the PGT, in particular the possible physics
of the dynamic spin-0 modes; (2) to consider the prospects of
accounting for the outstanding present day mystery---the
accelerating universe---in terms of an alternative gravity theory,
more particularly in terms of the PGT. With the usual assumptions of
isotropy and homogeneity in cosmology, we find that, under the
model, the Universe will oscillate with generic choices of the
parameters. The $0^+$ dynamic mode in the model plays the role of
the imperceptible ``dark energy''. With a certain range of parameter
choices, it can account for the current status of the Universe,
i.e., an accelerating expanding universe with a value of the Hubble
constant which is approximately the present one. These promising
results should encourage further investigations of this model, along
with a detailed comparison of its predictions with the observational
data.

The remainder of this work is organized as follows: We summarize the
formulation of the PGT in general and our model with scalar and
pseudoscalar modes in Sec.~2, and then consider the PGT scalar mode
cosmological model in Section~3.  In Section~4 an effective
Lagrangian and Hamiltonian for our cosmological model is presented.
This is followed by a late-time asymptotic expansion in Section~4 in
which certain normal modes are identified.  Section~6 includes the
results of our numerical demonstrations for various choices of the
parameters and the initial data along with a comparison with the
supernova observations. The implications of our findings are
discussed in Section~7 and Sec.~8 is a conclusion.

Throughout the paper our conventions are as follows: The spacetime
signature is $(-, +, +, +)$ and $c=\hbar=1$. The Greek indices,
$\alpha,\beta,\gamma\dots $, are 4d orthonormal (an-holonomic)
indices, whereas the Latin indices $i,j,k\dots $ are 4d coordinate
(holonomic) indices; they both range over $0,1,2,3$. On the other
hand the Latin indices $a,b,c,d$ are 3-dimensional, with range
$1,2,3$.










\section{The Poincar\'e Gauge Theory}

In the Poincar\'e gauge theory of gravity (PGT) \cite{Hehl80,HS80},
the two sets of local gauge potentials are, for ``translations'',
the orthonormal co-frame $\vartheta^\alpha=e^\alpha{}_i {\rm d}x^i$,
where the metric is
$g=-\vartheta^0\otimes\vartheta^0+\delta_{ab}\vartheta^a\otimes\vartheta^b$,
and, for ``rotations'', the metric-compatible (Lorentz Lie-algebra
valued) connection 1-forms
$\Gamma^{\alpha\beta}=\Gamma^{[\alpha\beta]}{}_i {\rm d}x^i$. The
associated field strengths are the torsion and curvature 2-forms
\begin{eqnarray}
T^\alpha&:=&{\rm d}\vartheta^\alpha+\Gamma^\alpha{}_\beta\wedge
\vartheta^\beta=\frac12 T^\alpha{}_{\mu\nu}\vartheta^\mu\wedge\vartheta^\nu,
\label{torsion}\\
R^{\alpha\beta}&:=&{\rm d}\Gamma^{\alpha\beta}
+\Gamma^\alpha{}_\gamma\wedge\Gamma^{\gamma\beta}=
\frac12R^{\alpha\beta}{}_{\mu\nu}\vartheta^\mu\wedge\vartheta^\nu,
\label{curvature}
\end{eqnarray}
which satisfy the respective Bianchi identities:
\begin{equation}
DT^\alpha\equiv R^\alpha{}_\beta\wedge \vartheta^\beta,\qquad
DR^\alpha{}_\beta\equiv0.\label{bianchi}
\end{equation}

The  PGT Lagrangian density is taken to have the standard quadratic
form; qualitatively,
\begin{equation}
{\mathscr L}[\vartheta,\Gamma]\sim\Lambda - a_0 R
+\sum_{{\mathsf{n}}=1}^3 a_{\mathsf{n}}{\buildrel{(\mathsf{n})}\over
T}{}^2+\sum_{\mathsf{n}=1}^6 b_\mathsf{n}
{\buildrel{(\mathsf{n})}\over R}{}^2,
\end{equation}
where $\displaystyle{\buildrel{(\mathsf{n})}\over T}$ and
$\displaystyle{\buildrel{(\mathsf{n})}\over R}$ are the
algebraically irreducible parts of the torsion and curvature and
$\Lambda$ is the cosmological constant.  
 The
gravitational field equations obtained from varying with respect to
the respective gauge potentials $\vartheta^\alpha{}_i$,
$\Gamma^{\alpha\beta}{}_j$ have the qualitative form
\begin{eqnarray}
\fl\quad\Lambda - a_0 G_\alpha{}^i+ \sum_{\mathsf{n}=1}^3
a_{\mathsf{n}} ( D {\buildrel{(\mathsf{n})}\over T}
+{\buildrel{(\mathsf{n})}\over T}{}^2 ) +\sum_{\mathsf{n}=1}^6
b_{\mathsf{n}} {\buildrel{(\mathsf{n})}\over R}{}^2&\sim&
\hbox{source energy-momentum density},\\
\quad\,a_0 T+\sum_{\mathsf{n}=1}^3 a_{\mathsf{n}}
{\buildrel{(\mathsf{n})}\over T}+ \sum_{\mathsf{n}=1}^6
b_{\mathsf{n}} D{\buildrel{(\mathsf{n})}\over R}&\sim&\hbox{source
spin density}.
\end{eqnarray}
These are, respectively, second order equations\footnote{It should
be noted that the PGT is obtained from an action containing
quadratic curvature terms, but with the connection as a variable
they yield 2nd order equations.  Higher order equations {\em would}
result from such an action {\em if} one used the Christoffel
connection---or decomposed the Lorentz connection into a Christoffel
part plus torsion terms and then treated the torsion and the metric
as the fundamental dynamical fields.  Such a decomposition is alien
to gauge principles.} for $\vartheta^\alpha{}_i$ and
$\Gamma^{\alpha\beta}{}_j$. In conjunction with the Bianchi
identities (\ref{bianchi}), these two equations yield, respectively,
the conservation of source energy-momentum and angular momentum
statements.


Here, generalizing an earlier work \cite{SNY08}, we wish to examine
the dynamics of the special case describing the  two good PGT
dynamic scalar modes in a cosmological model.  For this two spin-0
modes case we should take $b_{\mathsf{n}}=0$ except for $b_6\to b^+$
and $b_3 \to -b^-$ (see Appendix A). For more convenient signs we
also make the replacement $a_{\mathsf{n}}\to -A_{\mathsf{n}}$.

The gravitational Lagrangian of our model has the specific form
\begin{equation}
{\mathscr L}[\vartheta,\Gamma]=\frac{1}{2\kappa}\left[- 2 \Lambda +
A_0 R -\frac12\sum_{\mathsf{n}=1}^3 A_{\mathsf{n}}
{\buildrel{(\mathsf{n})}\over
T}{}^2+\frac{b^+}{12}R^2+\frac{b^-}{12}E^2\right], \label{Ldensity}
\end{equation}
where $\kappa=8\pi G$, $R$ is the scalar curvature and $E$ is the
pseudoscalar curvature (specifically $E/6=R_{[0123]}$ is the
magnitude of the one component of the totally antisymmetric
curvature). The cosmological constant has been included both for
generality and a comparison with other models.

In detail, the first field equation, obtained from variation with
respect to the orthonormal frame, has the components
\begin{eqnarray}
\fl&&\Lambda g_{\mu\nu} + A_0 G_{\mu\nu} + A_1 \left(\nabla_\alpha
T_{\nu\mu}{^\alpha} + \frac{1}{2} T_{\nu\alpha\beta}
T_\mu{^{\alpha\beta}} - T_{\nu\mu\alpha} T^\alpha + \frac{1}{4}
g_{\mu\nu}  T^{\alpha\beta\gamma} T_{\alpha\beta\gamma} -
T_{\alpha\beta\mu} T^{\alpha\beta}{}_\nu \right)
\nonumber\\
\fl&&+ \frac{A_2-A_1}{3} \left(g_{\mu\nu}\nabla_\alpha
T^\alpha-\nabla_\nu
T_\mu - \frac{1}{2}g_{\mu\nu}T_\alpha T^\alpha\right) + \frac{A_3 - A_1}{18}
( 6 \nabla_\alpha P^\beta \epsilon^\alpha{_{\beta\nu\mu}} - 4 P_\alpha
T_{\nu\beta\gamma}\epsilon^{\beta\gamma\alpha}{_\mu} \nonumber\\
\fl&&+ 3 P_\alpha T^\alpha{_{\beta\gamma}}
\epsilon^{\beta\gamma}{_{\nu\mu}} - 4 T^\alpha{_{\beta\nu}}
P^{\gamma} \epsilon_\alpha{^\beta}{_{\gamma\mu}} - P_\alpha P^\alpha
g_{\mu\nu} + 2 P_\mu P_\nu ) - \frac{b^-}{24} (E^2 g_{\mu\nu} - 2 E
R_{\alpha\beta\gamma\nu}
\epsilon^{\alpha\beta\gamma}{_\mu}) \nonumber\\
\fl&&+\frac{b^+}{24}(4 R_{\mu\nu} - R g_{\mu\nu}) R = \kappa
{\mathscr T}_{\mu\nu}\label{1st}
\end{eqnarray}
where $ G_{\mu\nu}= R_{\mu\nu} - \frac{1}{2} g_{\mu\nu} R $ is the
Einstein tensor and ${\mathscr T}_{\mu\nu}$ is the (in general
nonsymmetric) material energy-momentum density tensor.

The components of the second field equation, obtained from the
variation with respect to the connection, can be decomposed into
three algebraically irreducible parts:
\begin{eqnarray}
(6 (A_0-A_1)+{b^+}R){\buildrel{(1)}\over T}{^\alpha}{_{\beta\gamma}}
- \frac{b^-}{2}E {\buildrel{(1)}\over
T}{^\alpha}{}_{\mu\nu}\epsilon^{\mu\nu}{}_{\beta\gamma}
&=&0,\label{T1}\\
b^+\nabla_\mu R-\frac23(6m^{+}+{b^+}R)T_\mu + \frac13 b^{-}EP_\mu&=&0,
\label{T2}\\
b^-\nabla_\mu E - \frac13(6m^{-}+b^{+}R)P_\mu-\frac23
b^-ET_\mu&=&\kappa S_\mu=0,\label{T3}
\end{eqnarray}
where $T_\mu\equiv T^\alpha{}_{\alpha\mu}= {\buildrel{(2)}\over
T}{}^\alpha{}_{\alpha\mu}$,
$P_\mu\equiv\frac{1}{2}\epsilon_{\mu\nu}{}^{\alpha\beta}T^\nu{}_{\alpha\beta}
=\frac{1}{2}\epsilon_{\mu\nu}{}^{\alpha\beta} {\buildrel{(3)}\over
T}{}^\nu{}_{\alpha\beta}$ are the torsion trace and axial vectors,
$m^{+}\equiv A_0+A_2/2$ and $m^{-}\equiv A_0+2A_3$ are the masses of
the respective linearized modes, and $S_\mu$ is the axial vector
spin density which, for simplicity, we have assumed to vanish
 (this should be a good approximation except at high densities
such as those expected in the very early universe). The $0^-$ part
couples, as indicated here, to the axial spin vector of spin-1/2
fermions, but the $0^+$ mode {\em does not couple to any known
source}.

From (\ref{T1}) we find the general solution ${\buildrel{(1)}\over
T}{}^\alpha{}_{\beta\gamma}=0$. From (\ref{T2}) and (\ref{T3}) we
learn that the torsion trace and axial vectors are controlled by the
gradients of two functions. This reflects their respective spin
$0^+$, spin $0^-$ fundamental nature. However, in view of the
non-linearities of these relations as well as the geometric
significance of the respective ``potential functions'' (i.e., they
are the scalar and pseudoscalar curvatures), one can see that it is
neither possible nor appropriate to resolve them to find new,
simpler dynamics for two ``scalar potentials''.

\section{The PGT scalar mode cosmological model}
For a homogeneous, isotropic  FLRW
(Friedmann-Lema\^{i}tre-Robinson-Walker) cosmological model the
isotropic orthonormal coframe has the form
\begin{equation}
\vartheta^0={\rm d}t,\qquad \vartheta^a=a(t)\,(1+ \frac{1}{4}k
r^2)^{-1}\,{\rm d}x^a,
\end{equation}
where $k=-1,0,+1$ is the sign of the Riemannian spatial curvature.
Here we will consider for simplicity only the flat $k=0$ case (as
far as the observations can tell, this appears to well describe our
physical universe).

Because of isotropy, for this $k=0$ case the only non-vanishing 
connection one-form coefficients are of the form \begin{equation} \Gamma^a{}_0=\Psi(t)\,
{\rm d}x^a,\qquad \Gamma^a{}_b=X(t)\epsilon^a{}_{bc}\, {\rm d}x^c,
\end{equation}
where $\epsilon_{abc}:= \epsilon_{0abc}$ is the usual 3 dimensional
Levi-Civita anti-symmetric tensor. From the definition of the
curvature (\ref{curvature}), one can now find all the nonvanishing
curvature 2-forms:
\begin{equation}
\fl\quad R^{0a}=\dot\Psi\, {\rm d}t\wedge{\rm
d}x^a-X\Psi\epsilon^a{}_{bc}{\rm d}x^b\wedge{\rm d}x^c,\quad
R^{ab}=\dot X\epsilon^{ab}{}_c\, {\rm d}t\wedge{\rm
d}x^c+(\Psi^2-X^2){\rm d}x^a\wedge{\rm d}x^b.
\end{equation}
%
Consequently, the scalar and pseudoscalar curvatures are,
respectively,
\begin{eqnarray}
R&=& 6[a^{-1}\dot\Psi + a^{-2}(\Psi^2- X^2)], \label{R} \\
E&=& 6[a^{-1}\dot{X} +  2 a^{-2}X\Psi].\label{E}
\end{eqnarray}

Because of isotropy, the only nonvanishing  torsion tensor
components are of the form
\begin{equation}
T^a{}_{b0}=f(t)\delta^a_b, \qquad T^a{}_{bc}=-
2\chi(t)\epsilon^a{}_{bc}.
\end{equation}
From the definition of the torsion (\ref{torsion}) one can find the
relation between the torsion components and the gauge variables:
\begin{equation}
f=a^{-1}(\Psi-\dot a), \qquad \chi= a^{-1}X. \label{fchi}
\end{equation}
(Note: the variable $\Phi=-3f$ was used in the earlier work
\cite{SNY08}.)

From the isotropic assumption, the material energy momentum tensor
must have the perfect fluid form.  In this work we focus on the late
time behavior. Accordingly we assume that the fluid pressure can be
neglected, so that the gravitating material behaves like dust with a
density satisfying $\rho a^3=\hbox{constant}$. Also, we remind the
reader that, although we expect the spin density to play an
important role in the early universe, it is reasonable to assume
that the material spin density is negligible at late times.

Due to isotropy, the first field equation (\ref{1st}) has only two
nontrivial distinct components.  Expressed in terms of the tensorial
quantities and the Hubble parameter $H=\dot a/a$ they are the ``00''
piece
\begin{eqnarray}
\fl\quad&&+\Lambda+\frac{3A_2}{2}H^2-3m^+(H+f)^2+3m^-\chi^2\nonumber\\
 \fl\quad&&\qquad+\frac{b^-}{24}E^2 - b^-E(H+f)\chi+ \frac{b^+}{24}R^2
 -\frac{b^+}{2}R [(H+f)^2 - \chi^2] =-\kappa \rho,\label{G00}
\end{eqnarray}
which contains only first time derivatives of the potentials (and is
hence an initial value constraint), and the ``space-space'' piece
\begin{eqnarray}
\fl\quad&&-\Lambda + \frac{m^+}{3} [R-3(H+f)^2] + (2m^+-m^-) \chi^2
-\frac{A_2}{2} (2 \dot{H} + 3 H^2)\nonumber\\
\fl\quad&&\qquad+\frac{b^-}{72} E^2 - \frac{b^-}{3} E (H+f)\chi +
\frac{b^+}{72} R^2 -  \frac{b^+}{6} R[(H+f)^2 - \chi^2] = -\kappa
p=0, \label{Gkk}
\end{eqnarray}
 a dynamical equation for $\ddot a$.

From the components of the second field equation
(\ref{T2},\ref{T3}), we obtain
\begin{eqnarray}
b^+\dot R&=&2(b^+R+6m^+)f+2b^-E\chi,\label{ddtf}\\
b^-\dot E&=&2b^- E f-2 (b^+R+6m^-)\chi,\label{ddtx}
\end{eqnarray}
which---along with the definition of the curvature scalars
(\ref{R}), (\ref{E}) --- are second order dynamical equations for the
connection coefficients.

Before we present our more detailed discussion of these dynamical
equations equations it should be noted that there are some special
``non-dynamic effective cosmological constant'' cases---that is
special cases having one constant magnitude field when certain
coefficients a/o other field components vanish. Such field
components would contribute to the dynamical equations certain
constant terms that would act like effective cosmological constants.
Since here we are interested in the generic case with both modes
dynamic, we only mention these special cases briefly and cite other
works where they have been considered in more detail.  In particular
\cite{SNY08} discussed the case with vanishing $b^-$, $\chi$ and
$R=-6m^+/b^+$, while \cite{MGK07} has considered the case with
vanishing $b^-$, $f$, and $R=-6m^-/b^+$. It should also be mentioned
that the earlier investigators have tended to decompose the
connection into its Christoffel part plus some torsion.  Using such
a decomposition leads to higher order equations---unless one takes
$A_2=0$, as these investigators were prone to do (see, e.g.,
\cite{Min80,Min83,MN95,MGK07,GMH}. Here {\em we do not
make such a decomposition of the connection} and need not make such
a parameter restriction---which would in fact render the $0^+$ mode
non-dynamic.


Instead of considering the three second-order differential equations
(\ref{Gkk},\ref{ddtf},\ref{ddtx},), we can transform these
second-order equations into six first-order differential equations
by including the definition of $H$ and (\ref{R})--(\ref{E}) for the
six unknown variables: $a$, $H$, $f$, $\chi$, $R$, and $E$ (where
all except $a$ and $H$ are gauge covariant tensor fields).  For some
purposes this is more convenient; in particular first-order
differential equations are more suitable for numerical calculations.
Combining Eqs.~(\ref{G00}) and (\ref{Gkk}) gives
\begin{equation}
-3A_2(\dot{H}+2H^2)+m^{+}R-4\Lambda
+6(m^{+}-m^{-})\chi^2=\kappa(\rho -3p)=\kappa\rho\label{3Gkk-G00}.
\end{equation}
Eq.~(\ref{3Gkk-G00}) is considered as a first-order differential
equation for $H$. Moreover, Eqs.~(\ref{ddtf}) and (\ref{ddtx}) are
already first-order differential equations for $R$ and $E$. After
replacing $\dot{H}$ in Eq.~(\ref{R}) by using (\ref{3Gkk-G00}), it
is clear that Eqs.~(\ref{R}) and (\ref{E}) give the first-order
equations for $f$ and $\chi$. With a straightforward
re-organization, the six equations are
\begin{eqnarray}
\dot{a} &=& aH, \label{dta}\\
\dot{H} &=&\frac{m^+}{3 A_2} R+\frac{2 (m^{+}-m^{-})}{ A_2} \chi^2-2H^2
-\frac{\kappa\rho}{3 A_2}-\frac{4\Lambda}{3A_2},\label{dtH}\\
\dot{f} &=&\frac{4A_3}{A_2}\chi^2-\frac{A_0}{3A_2}R-f^2-3Hf
+\frac{\kappa\rho}{3 A_2}+\frac{4\Lambda}{3 A_2}, \label{dtf}\\
\dot{\chi} &=& \frac{E}{6}-(3H+2f)\chi,\label{dtchi}\\
\dot{R}&=&\frac{ 2 b^-}{b^+}\chi E+2f\left(R+\frac{6 m^+}{b^+}\right),
\label{dtR}\\
\dot{E}&=&2fE-\frac{2 b^+}{b^-}\chi\left(R+\frac{6m^-}{b^+}\right),\label{dtE}
\end{eqnarray}
with the constraint equation
\begin{eqnarray}
&&\Lambda + \frac{3}{2}A_2 H^2 + 3 m^{-}\chi^2-3m^{+}(f+H)^2
+\frac{b^+}{24}R^2+\frac{b^-}{24} E^2\nonumber\\
&&\qquad\qquad\quad-b^- E (f+H)\chi-\frac{b^+}{2} R [(f+H)^2-\chi^2]
= -\kappa\rho.\label{firsteq_7}\label{constraint}
\end{eqnarray}
The constraint equation can be used to replace $\kappa\rho$ in
Eqs.~(\ref{dtH},\ref{dtf}) to give alternative versions of these two
equations:
\begin{eqnarray}\fl\qquad
\dot H&=&-\frac1{A_2}\Lambda+\frac{m^+}{3A_2}R-\frac32
H^2+\frac{2m^+-m^-}{A_2}\chi^2-\frac{m^+}{A_2}(f+H)^2\nonumber \\
\fl\qquad&&+\frac{1}{3A_2}\left\{\frac{b^+}{24}R^2-\frac{b^+}{2}R
\left[(f+H)^2-\chi^2\right]+\frac{b^-}{24}E^2-b^-E
(f+H)\chi\right\}, \label{Hdotnosource}\\
\fl\qquad
 \dot f &=& \frac1{A_2}\Lambda-\frac{A_0}{3A_2}R-\frac12
f^2-2Hf+\frac{2A_3}{A_2}\chi^2+\frac{A_0}{A_2}[(f+H)^2-\chi^2]\nonumber \\
\fl\qquad&&-\frac{1}{3A_2}\left\{\frac{b^+}{24}R^2-\frac{b^+}{2}R
\left[(f+H)^2-\chi^2\right]+\frac{b^-}{24}E^2-b^-E
(f+H)\chi\right\}\label{fdotnosource1}.
\end{eqnarray}
These two alternative equations along with the four other first order
equations make a closed system for the geometric variables which is
more practical for numerical evolution. This alternative system will
be obtained in another way in the next section.

For a comparison with GR models, the ``00'' constraint (\ref{G00})
can be considered as a generalized Friedman equation:
\begin{equation}
3 A_0 H^2 = \kappa (\rho + \rho_{\Gamma})+\Lambda,
\label{energy_effective}
\end{equation}
where the effective energy due to the dynamic connection is
\begin{equation}
\fl\kappa\rho_{\Gamma}=3m^{-}\chi^2-6m^{+}Hf-3m^{+}f^2+\frac{b^+}{24}R^2
+\frac{b^-}{24}E^2-b^{-}E\chi(H+f)-\frac{b^+}{2}R[(H+f)^2-\chi^2],
\label{rhoGamma}
\end{equation}
Moreover, the ``space-space'' Eq.~(\ref{Gkk}) may be considered as a
force-balance equation:
\begin{equation}
 A_0\left(\frac{\ddot{a}}{a}\right)=-\frac{\kappa (\rho+3p)}{6}
 -\frac{\kappa(\rho_{\Gamma}+3p_{\Gamma})}{6}+\frac{\Lambda}{3},
\label{pressure_effective}
\end{equation}
where the effective pressure due to the dynamic connection is
\begin{eqnarray}
\kappa p_{\Gamma}&=&2m^{+}\dot{f}+m^{+}f(4H+f)-m^{-}\chi^2\nonumber\\
&+&\frac{b^+}{72}R^2+\frac{b^-}{72}E^2-\frac{b^-}{3}E\chi(H+f)
-\frac{b^+}{6}R[(H+f)^2-\chi^2].\label{pGamma}
\end{eqnarray}
It can here be seen how $\rho_{\Gamma}+3p_{\Gamma}<0$, which is
indeed possible, could produce an accelerated universe. Although
these relations are useful for comparison with other models, an
examination of the dynamical terms implicit in $\rho_{\Gamma}$ and
$p_{\Gamma}$ shows that their PGT dynamic nature differs
fundamentally from that of their GR counterparts.  In particular the
ratio $w_{\Gamma}:=P_{\Gamma}/\rho_{\Gamma}$ can take on any value
and should not be given the usual equation-of-state interpretation.

\section{Effective Lagrangian and Hamiltonian}

Our cosmological model system of ODEs resemble those of a particle
with three degrees of freedom.  One may suspect that they can be
obtained directly from a variational principle. To achieve such a
goal it is natural to consider imposing the homogeneous-isotropic
symmetry into the field theory Lagrangian density. We note that it
has long been known that imposing symmetries and variations do not
commute in general. However, for GR they are known to commute for
all Bianchi class A cosmologies \cite{AS91}.  We conjecture that this
is also true for the PGT. In particular our $k=0$ model is an
isotropic Bianchi I (class A) model, so there is good reason to be
hopeful. Here we show that (at least for our dust fluid model)
imposing the FLRW symmetry on the PGT Lagrangian does indeed lead us
to the same expressions as were found from imposing the symmetry on
the field equations.

Imposing the FLRW symmetry on the Lagrangian density
(\ref{Ldensity}) leads to the effective Lagrangian
\begin{equation}
L_{\mathrm{eff}}=\frac{a^3}{\kappa}\left[-\Lambda+\frac{A_0}{2}R
+\frac{3}{2}A_2 f^2 -
6A_3\chi^2+\frac{b^+}{24}R^2+\frac{b^-}{24}E^2\right].
\end{equation}
It should be noted that (for least action) the coefficients of the
quadratic kinetic terms ($f^2$, $R^2$, $E^2$) which contain the time
derivatives---see the specific expressions for the curvature and
torsion Eqs.~(\ref{R},\ref{E},\ref{fchi})---must be non-negative.

Now we use this effective Lagrangian (along with the just mentioned
expressions for the curvature and torsion) to obtain a conserved
energy and three second order equations for the gauge parameters:
the connection coefficients $\psi$, $X$ and the frame/metric scale
factor $a$ . We also note that the second order Lagrange equations
can be rearranged and combined with the formulas for curvature and
torsion to give exactly the six first order equations obtained from
the general 4D covariant field equations specialized to the $k=0$
FLRW geometry. Following this we will then find the associated
Hamiltonian equations.

Making use of the formulas for the torsion and curvature components
in terms of the gauge variables, (\ref{R},\ref{E},\ref{fchi}), the
conserved energy function associated with $L_{\mathrm{eff}}$ is
found to be
\begin{eqnarray}
\fl\ {\cal E}:=\dot{q}^k \frac{\partial
L_{\mathrm{eff}}}{\partial\dot{q}^k}-
L_{\mathrm{eff}}&=&\frac{a^3}{\kappa}\left\{\Lambda  +
\frac{3A_2}{2}H^2
 -3m^+(f+H)^2 + 3m^-\chi^2\right .\nonumber\\
&&+\left .\frac{b^-}{24}E^2 - b^-E(H+f)\chi
    + \frac{b^+}{24}R^2 - \frac{b^+}{2}R[(H+f)^2 -
    \chi^2]\right\},\label{energyfunction}
\end{eqnarray}
where $q_k=\{\Psi,X, a\}$. This is a combination that we recognize
from our earlier analysis, it is just the ``00 constraint''
(\ref{constraint}); its constant magnitude is the physical
combination $- a^3\rho$, which is indeed a constant because of the
dust fluid energy-momentum conservation relation.

Making use of the formulas for the torsion and curvature components
in terms of the gauge variables (\ref{R},\ref{E},\ref{fchi}) we now
obtain the Euler-Lagrange equations $\displaystyle{ \frac{d}{d t}
\frac{\partial L_{\mathrm{eff}}}{\partial \dot{q_i}} -
\frac{\partial L_{\mathrm{eff}}}{\partial q_i} =0}$.

For the $\Psi$ equation:
\begin{equation}
\fl\ \frac{d}{dt}\frac{\partial L_{\mathrm{eff}}}{\partial
\dot{\Psi}}
       = \frac{d}{dt}\left(3A_0a^2 + \frac {b^+}{2}a^2R\right) =
       \frac{\partial L_{\mathrm{eff}}}{\partial\Psi}
       = 6A_0a\Psi + b^+aR\Psi +b^- a EX + 3A_2a^2f.
\end{equation}
This is a second order equation for $\Psi$. (Here and below we
dropped for simplicity the overall factor of $\kappa^{-1}$.) It can
alternately be rearranged using (\ref{fchi}) to give (\ref{dtR}),
the first order equation for $\dot R$.

For the $X$ equation:
\begin{equation}
\fl\qquad \frac{d}{dt}\frac{\partial L_{\mathrm{eff}}}{\partial
\dot{X}}
       =\frac{d}{dt}\left(\frac{1}{2}b^-a^2E\right)=
\frac{\partial L_{\mathrm{eff}}}{\partial X}
       = b^-aE\Psi- b^+aX R -6A_0aX - 12A_3aX.
\end{equation}
This second order equation for $X$  can be rearranged using
(\ref{fchi}) into (\ref{dtE}), the first order equation for $\dot
E$.

For the $a$ equation:
\begin{eqnarray}
\fl\frac{d}{dt}\frac{\partial L_{\mathrm{eff}}}{\partial\dot{a}}
=\frac{d}{dt}(-3A_2a^2f) &=& \frac{\partial L_{\rm eff}}{\partial
a}=A_0[a^2R-3(\Psi^2-X^2)]+
\frac{b^+}{24}a^2R^2-\frac{b^+}{2}R(\Psi^2-X^2)\nonumber\\
&&\ \quad+\frac{b^-}{24}a^2E^2-b^-X\Psi
E+\frac{3}{2}A_2a^2f^2-6A_3X^2-3\Lambda a^2.
\end{eqnarray}
This is a second order equation for $a$.  It can be rearranged into
a first order equation for $\dot f$; the result
 is exactly (\ref{fdotnosource1}), the aforementioned alternative
to (\ref{dtf}) obtained by using (\ref{constraint}).
Now using (\ref{fchi}) one can calculate
\begin{equation}
\dot{H} = \frac{d}{dt}(a^{-1}\Psi)-\dot f=-H(H+f)+a^{-1}\dot
\Psi-\dot f,
\end{equation}
then using the just mentioned expression for $\dot f$ and (\ref{R})
one gets (\ref{Hdotnosource}). Moreover, from (\ref{E}) using
(\ref{fchi}) one gets the $\dot \chi$ equation (\ref{dtchi}).


It is remarkable that here in these Lagrange equations and the
associated conserved energy we get (at least for this dust case)
exactly the correct equations for our model---{\em without including
any explicit source coupling!}

We have cast our system into six first order equations for (3D)
tensorial quantities, equations which are suitable for numeric
evolution and comparison with observations. However these equations
are probably not in the most suitable form for the most penetrating
analytic analysis. So we here also present the Hamiltonian equations
for our PGT cosmology.

From the above one can introduce the canonical conjugate momentum
variables:
\begin{eqnarray}
P_a&\equiv&\frac{\partial L_{\rm eff}}{\partial \dot{a}} = - \frac{3 A_2}
{\kappa} a^2\, f ,\\
P_{\Psi}&\equiv&\frac{\partial L_{\rm eff}}{\partial \dot{\Psi}}
=a^2\left(\frac{3 A_0}{\kappa}  + \frac{b^+}{2 \kappa} R\right),\\
P_{X}&\equiv&\frac{\partial L_{\rm eff}}{\partial \dot{X}} =
\frac{b^-}{2\kappa} a^2 E.
\end{eqnarray}
Now one can construct the effective Hamiltonian:
\begin{eqnarray}
\fl\ H_{\rm eff}=P_a \dot{a} + P_{\Psi} \dot{\Psi} + P_{X} \dot{X} -
L_{\rm eff} &=&\frac{\kappa}{6 a}\left(\frac{P_a^2}{A_2} +
\frac{P_{X}^2}{b^-} + \frac{P_{\Psi}^2}{b^+}\right)-\left(\frac{A_0
a^2}{ b^+} +\Psi^2 - X^2\right)
\frac{P_{\Psi}}{a}\nonumber\\
&+&{\Psi} P_a - \frac{2}{a} X\, \Psi P_{X} +\frac{6 A_3}{\kappa} a
X^2 + \left[\frac{3 A_0^2}{2 b^+}+\Lambda
\right]\frac{a^3}{\kappa},\
\end{eqnarray}
 and obtain the six Hamilton equations:
\begin{eqnarray}
\dot{a}&=&\quad\frac{\partial H_{\rm eff}}{\partial P_a}=
\frac{\kappa P_a}{3A_2 a}+\Psi,\\
\dot{\Psi}&=&\quad\frac{\partial H_{\rm eff}}{\partial P_\Psi} =
\frac{\kappa}{3b^+ a}
P_\Psi-\frac{\Psi^2}{a}+\frac{X^2}{a}-\frac{A_0}{b^+}a,\\
\dot{X}&=&\quad\frac{\partial H_{\rm eff}}{\partial
P_X}=\frac{\kappa}{3b^- a}P_X
-\frac{2}{a}X\Psi,\\
\dot{P}_a&=&-\frac{\partial H_{\rm eff}}{\partial a}=\frac{H_{\rm
eff}-\Psi P_a}{a}
-\frac{12A_3}{\kappa}X^2-\left[\frac{2A_0^2}{b^+}+\Lambda\right]
\frac{3a^2}{\kappa}+\frac{2A_0}{b^+}P_\Psi,\\
\dot{P}_\Psi&=&-\frac{\partial H_{\rm eff}}{\partial\Psi}=
-P_a+\frac{2}{a}(\Psi P_\Psi+P_X X),\\
\dot{P}_X&=&-\frac{\partial H_{\rm eff}}{\partial X}=
-\frac{12A_3}{\kappa}aX +\frac{2}{a}(P_X\Psi-P_\Psi X).
\end{eqnarray}

This canonical reformulation should be of considerable interest for
further studies of this model, since the Hamiltonian formulation is
the framework for the most powerful known approaches for
analytically studying the dynamics of a system, including such
techniques as the Hamilton-Jacobi method and phase space portraits.
\section{Asymptotic Expansion}\label{asympexp}
  At late times in an expanding universe as the scale factor $a$
becomes larger the field amplitudes should be decreasing.  We can
then expect the quadratic terms in (\ref{energyfunction}) to be
dominant; hence, when the cosmological constant vanishes the late
time asymptotic behavior of $H$, $f$, $\chi$, $R$, and $E$ should
have a $a^{-3/2}$ fall off. So we reparametrize them according to
\begin{equation}
H=ha^{-3/2},\, f=ya^{-3/2},\, \chi=xa^{-3/2},\, R=ra^{-3/2},\, E=ea^{-3/2}.
\label{Asymp}
\end{equation}
The current Universe corresponds to $a^{3/2}\gg 1$.
Substituting (\ref{Asymp}) into (\ref{dta})--(\ref{firsteq_7}) gives
\begin{eqnarray}
\fl\qquad&& \dot{a}=a^{-1/2}h, \label{Asymp_1}\\
\fl\qquad&&\dot{h}=a^{-3/2}\left[\frac{2(m^+ -m^-)}{A_2}x^2-\frac{h^2}{2}
-\frac{\kappa}{3A_2}\rho_0 a^3(0)\right]+\frac{m^+}{3A_2}r
-a^{3/2}\frac{4\Lambda}{3A_2},\label{Asymp_2}\\ 
\fl\qquad&&\dot{y}=a^{-3/2}\left[\frac{4A_3}{A_2}x^2-y^2-\frac32 hy
+\frac{\kappa}{3A_2}\rho_0a^3(0)\right]-\frac{A_0}{3A_2}r+a^{3/2}
\frac{4\Lambda}{3A_2}, \label{Asymp_3}\\
\fl\qquad&&\dot{r}=a^{-3/2}\left[2yr+\frac32 rh+\frac{2
b^-}{b^+}ex\right]
+\frac{12 m^+}{b^+}y,\label{Asymp_4}\\
\fl\qquad&&\dot{x}=a^{-3/2}\left[-2xy-\frac{3}{2}hx\right]+\frac{e}{6},\\
\fl\qquad&&\dot{e}=a^{-3/2}\left[2ey+\frac32 he-\frac{2b^+}{b^-}rx\right]
-\frac{12m^-}{b^-}x,\label{Asymp_6}\\
\fl\qquad&&-a^3\kappa\rho=-\kappa\rho_0 a^3(0)=a^3\Lambda+\frac{3A_2}{2}{h}^2
 +3m^{-}{x}^2+\frac{b^-}{24}e^2+\frac{b^+}{24}r^2-3m^+ (y+h)^2\nonumber\\
\fl\qquad&&\qquad\qquad\qquad\qquad\quad-a^{-3/2}\left[b^- ex(h+y)
+\frac{b^+}{2}r((h+y)^2-x^2)\right].
\end{eqnarray}
Now let us restrict our further considerations to the vanishing
cosmological constant case. With
 $a^{3/2}\gg 1$, Eqs.~(\ref{Asymp_1})--(\ref{Asymp_6}) show that the
spin-$0^+$ mode and the spin-$0^-$ mode are becoming decoupled and
the asymptotic Hubble rate $h$ is influenced purely by the
spin-$0^+$ mode. Dropping the higher order terms (with $\Lambda=0$),
gives the energy constraint
\begin{equation}
\fl\qquad\qquad-a^3\kappa\rho=-\kappa\rho_0a^3(0)=\frac{3A_2}{2}{h}^2+3m^{-}x^2
+\frac{b^-}{24}e^2+\frac{b^+}{24}r^2-3m^+ (y+h)^2
\end{equation}
and three natural pairs of linear equations:
\begin{equation}\fl
(\dot{a}=a^{-1/2}h,\ \ \dot{h}=\frac{m^+}{3A_2}r),\quad
(\dot{y}=-\frac{A_0}{3A_2}r,\ \ \dot{r}=\frac{12m^+}{b^+}y),\quad
(\dot{x}=\frac{{e}}{6},\ \ \dot{e}=-\frac{12m^-}{b^-}x).
\end{equation}
The last two pairs, $(y,r)$ and $(x,e)$, are clearly harmonic
oscillators.  To analyze these equations further along with the
first pair, we introduce the new variable combination
\begin{equation}
z:= m^+y+A_0h.
\end{equation}
We then find three late time normal modes:
\begin{eqnarray}
\ddot{x}+\omega_-^2x&=&0,\qquad\mbox{where}\quad\omega_-^2=\frac{2m^-}{b^-}\\
\ddot{y}+\omega_+^2y&=&0,\qquad\mbox{where}\quad
\omega_+^2=\frac{4A_0m^+}{A_2b^+}\\
\dot{z}&=&0.
\end{eqnarray}
Reexpressed in terms of the late time normal modes the late-time energy
constraint is
\begin{equation}
\fl\qquad\quad\mathrm{const.}=-a^3\kappa\rho=-\frac3{A_0}z^2
+\left(\frac{3A_2}{2A_0}m^+y^2+\frac{b^+}{24}r^2\right)+
\left(3m^{-}x^2+\frac{b^-}{24}e^2\right),
\end{equation}
where each bracket is constant.
The physical and geometric significance of two of the normal modes
is clear, since they directly correspond to the two torsion
magnitudes (alternately the associated curvature scalars). Thus for
this model at late time we find that the two {\em dynamical
connection} modes are essentially {\em dynamical torsion modes} and
the description ``torsion cosmology'' is phenomenologically
appropriate, although it could lead to a misapprehension as to the
true fundamental dynamical fields.\footnote{ Although the
terminology ``torsion cosmology'' has often been used in the past to
describe the sort of model we are considering here, it has recently
come to our attention that such terms are fundamentally
inappropriate and can inhibit a deeper understanding.  The basic
dynamic variables in this theory are the orthonormal frame and
connection. }

The remaining mode $z$ corresponds to a certain combination of the
frame/metric scale expansion factor and the $0^+$ torsion:
\begin{equation}
Z=m^+f+A_0H
\end{equation}
which evolves according to
\begin{equation}
\dot{Z}=m^-\chi^2-m^+f(f+3H)-2A_0H^2+\frac{\kappa\rho}{6}.
\end{equation}
(With a nonvanishing cosmological constant this equation would pick
up an extra $2\Lambda/3$ term.) From the above we see that at late
time (with vanishing cosmological constant) the Hubble expansion
rate has the form
\begin{equation}
H=a^{-3/2}\times\mathrm{const.}-\frac{m^+}{A_0}f,
\end{equation}
with the $0^+$ torsion amplitude $f$ oscillating at the frequency
$\omega_+$.  Only the $0^+$ mode effects the expansion rate at late
times.  The late time acceleration is
\begin{equation}
\ddot a= a^{-1/2}\frac{m^+}{3A_2}r,
\end{equation}
which has periodic oscillations at the rate $\omega_+$. In this
model sometimes the expansion rate is accelerating and sometimes it
is slowing down.
\subsection{Numerical test}
The validity of our late time analytic results has been tested
numerically.  Taking the parameters as $A_0=1$, $A_2=0.23$,
$A_3=-0.35$, $b^{+}=1.1$, and $b^{-}=0.3$, we find $\omega_{+}=4.20$
and $\omega_{-}=1.4$. They have the relation
$\omega_{+}=3\omega_{-}$.
 Using the same parameters, we plot a full and a linear asymptotic
normal mode evolution of all the 6 dynamical equations.   The
behavior of the normal mode equations has been observed with several
sets of initial values.

\begin{figure}[thbp]
\begin{tabular}{rl}
\includegraphics[width=0.47\textwidth]{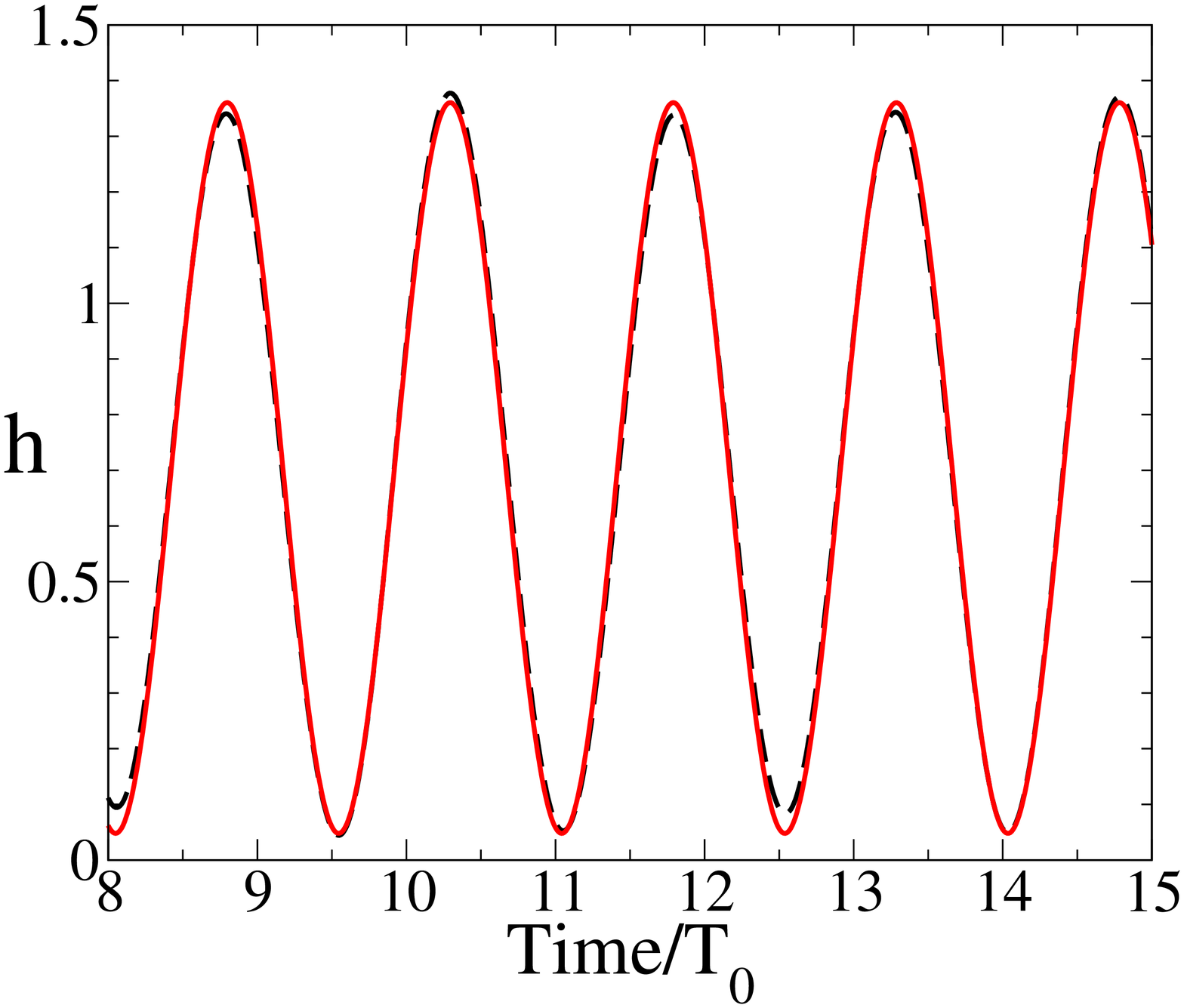}&
\includegraphics[width=0.47\textwidth]{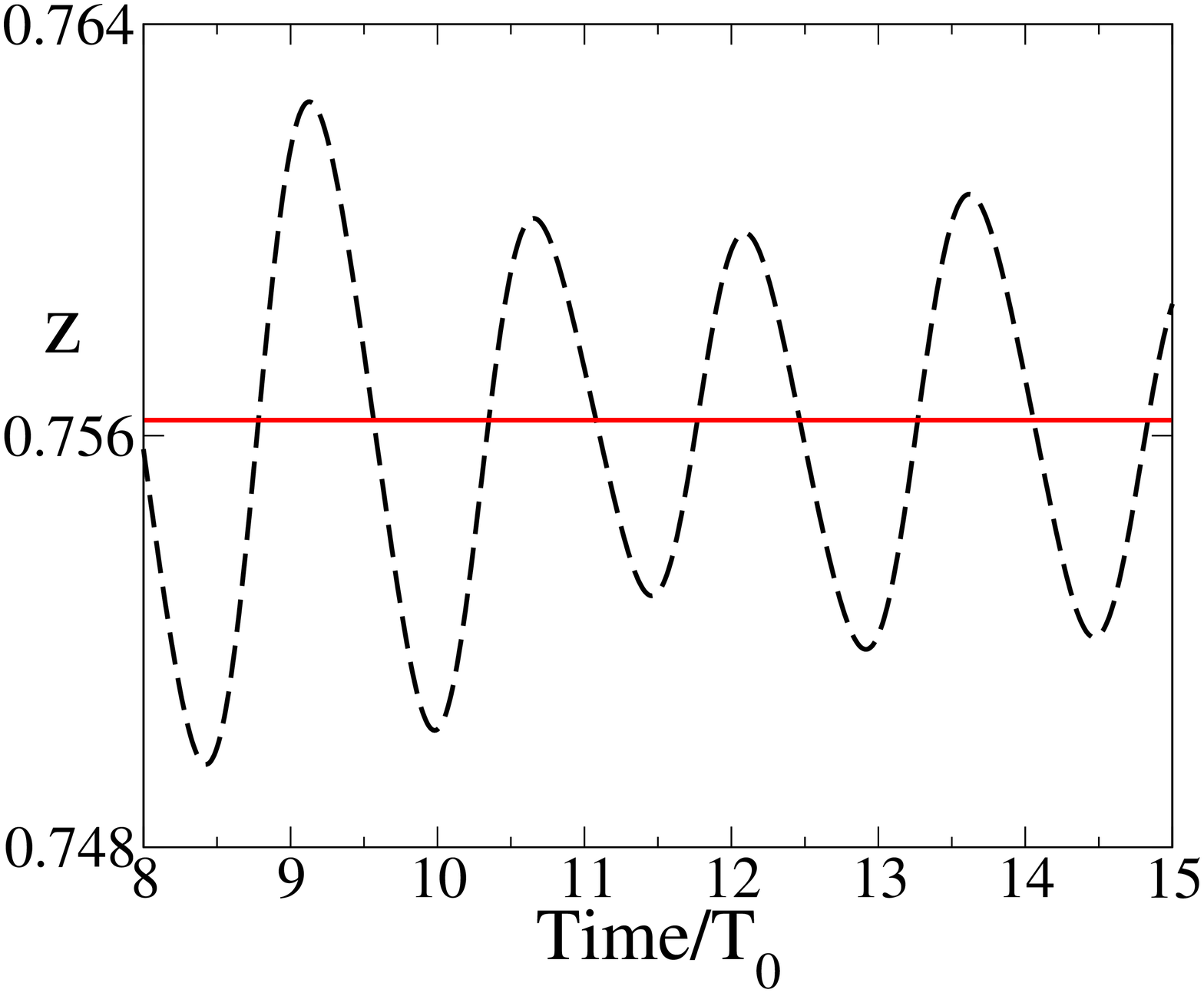}\\
\includegraphics[width=0.47\textwidth]{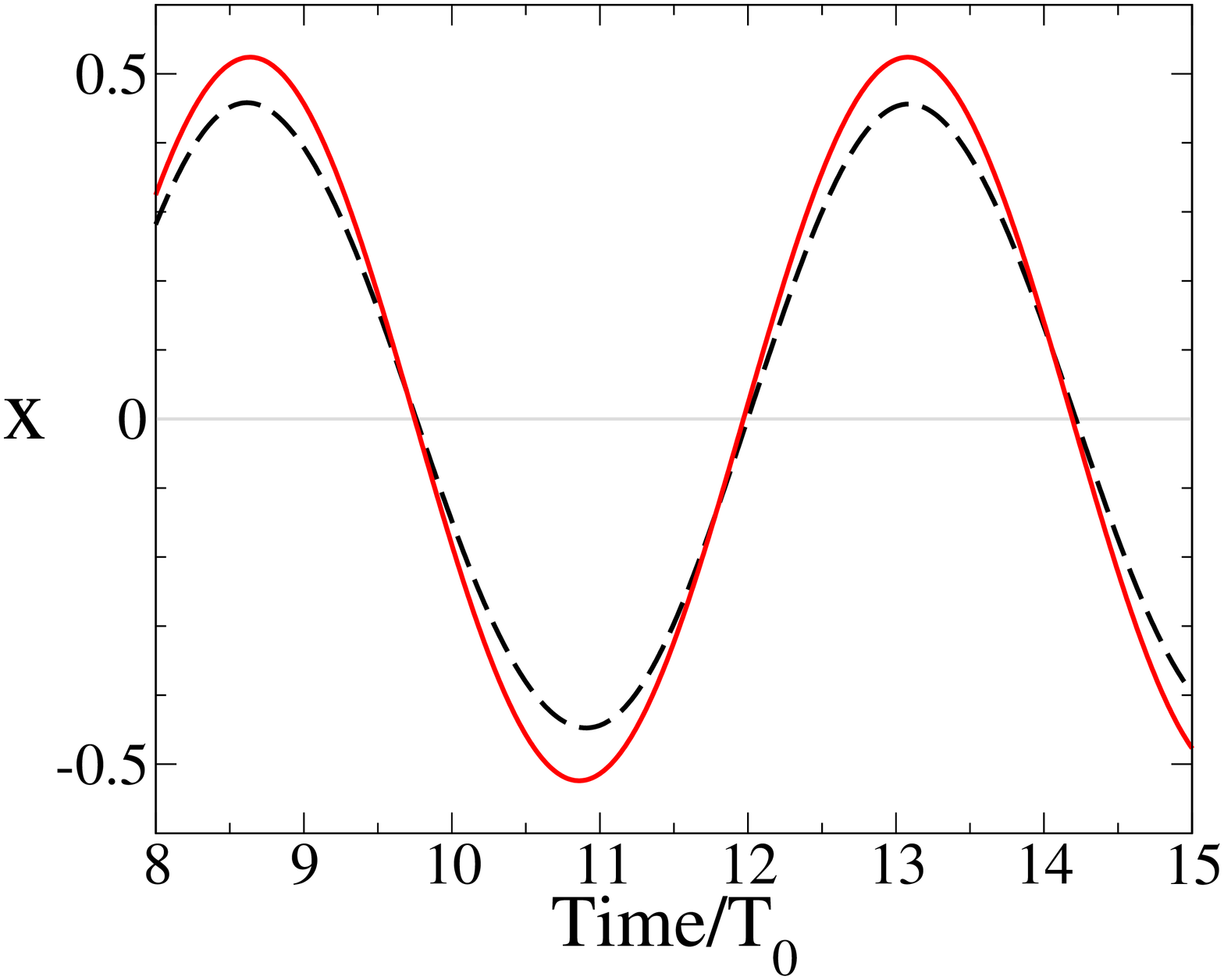}&
\includegraphics[width=0.47\textwidth]{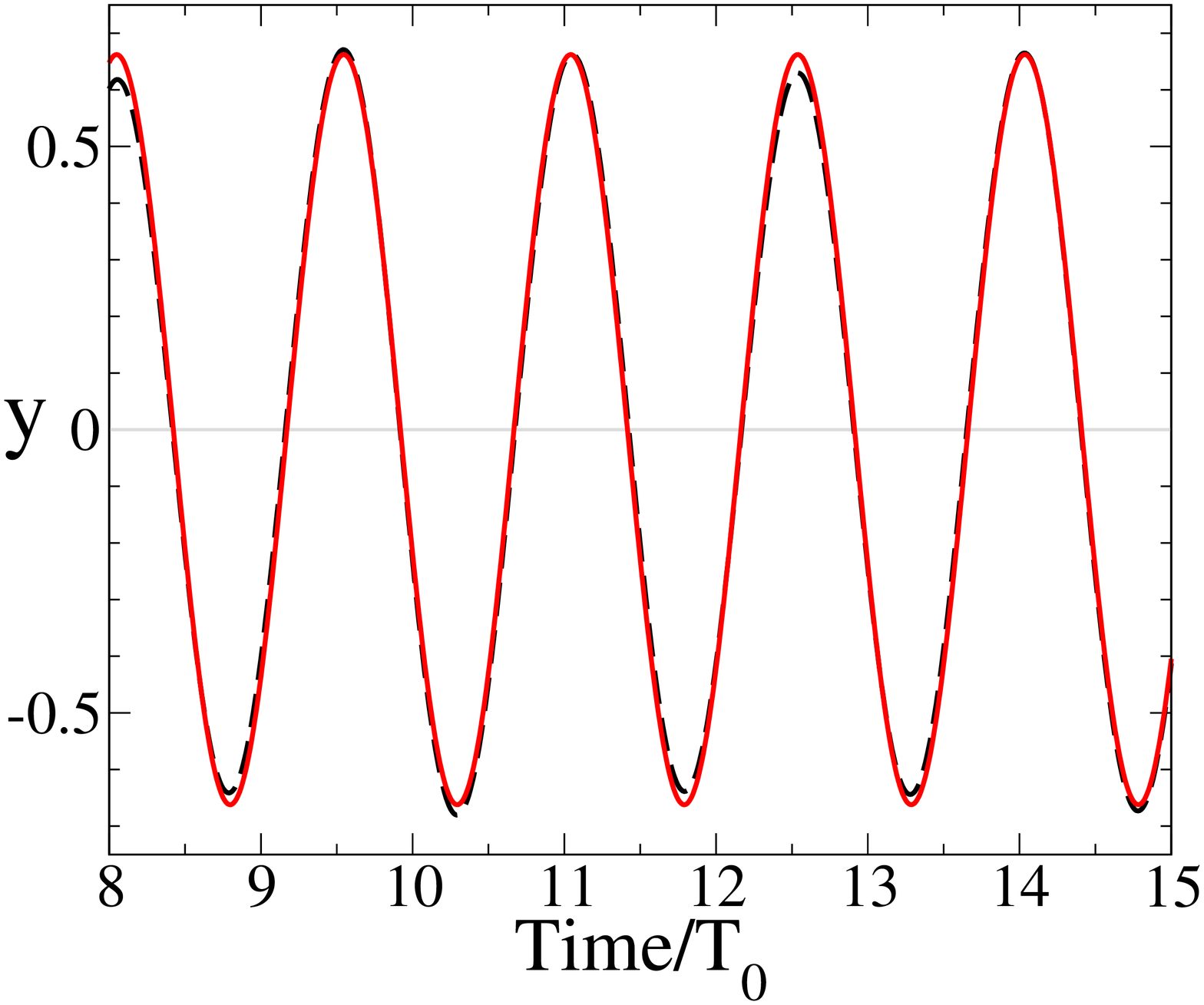}
\end{tabular}
\caption{Asymptotic evolution of the Hubble function, h, the late
time normal mode combination of torsion and Hubble function, $z$
(note the scale), the late time spin-$0^{-}$ normal mode, $x$, and
the late time spin-$0^{+}$ normal mode, $y$. The (black) dashed
lines represent the actual late time asymptotic evolution and the
(red) solid lines represent the linear approximation normal
modes.}\label{LA1}
\end{figure}

We have plotted one typical case in Fig.~\ref{LA1}. Here we show at
late time the asymptotic amplitudes: first, the Hubble function, h,
second, the normal mode combination of the torsion and Hubble
function, $z$, third, the spin-$0^{-}$ normal mode, $x$, fourth, the
spin-$0^{+}$ normal mode, $y$. The (black) dashed lines represent
the exact evolution and the (red) solid lines represent the late
time asymptotic normal mode behavior.

As expected, we found that the late time equations are indeed a good
approximation.  The $z(t)$ function approaches a flat line at late
time. The plots show that the frequency and amplitude of the full
and the linear approximation $0^+$ and $0^-$ asymptotic equations
are very close, although there are apparently still some nonlinear
effects.

\section{Numerical Demonstration}
In this section, we present the results of a numerical evolution of
our cosmological model. For all these calculations we take
$\Lambda=0$.  Since there is one scalar mode and one pseudoscalar
mode in this model, it is natural to investigate the interaction
between these two modes. We find that the pseudoscalar
connection mode can generate the scalar connection mode, but not
conversely. This offers a reason for the existence of the scalar
mode, since it is believed that the pseudoscalar mode exists and
plays an important role in the early universe due to its direct
interaction with matter. In Section \ref{preuni}, we first extend
the numerical demonstration of the earlier work \cite{SNY08} by
including the spin-$0^-$ mode, and then compare our numerical
results with the observational supernovae data. Not surprisingly we
find that the supernovae data can be better fitted with the
two-scalar-connection-mode model.

We need to look into the scaling features of this model before we
can obtain the sort of evolution results we seek on a cosmological
scale. In terms of fundamental units we can scale the variables and
the parameters as
\begin{eqnarray}
\fl&&t\rightarrow t/\ell,\quad a\rightarrow a,\quad H\rightarrow\ell
H,\quad f\rightarrow\ell f,\quad \chi\rightarrow\ell\chi,\quad
R\rightarrow\ell^2R,\quad E\rightarrow\ell^2 E,\nonumber\\
\fl&&A_0\rightarrow A_0,\quad A_2\rightarrow A_2,\quad
A_3\rightarrow A_3,\quad b^{+}\rightarrow b^{+}/\ell^2,\quad
b^{-}\rightarrow b^{-}/\ell^2,\quad
\Lambda\rightarrow\ell^2\Lambda,\label{scale1}
\end{eqnarray}
where $\ell^2\equiv\kappa=8\pi G$. So the variables and the scaled
parameters $b^+$ and $b^-$ become dimensionless (from the Newtonian
limit $A_0=1$).  Equations.~(\ref{dta}--\ref{dtE}) remain unchanged under
such a scaling. However, as we are interested in the cosmological
scale, it is practical to use another scaling---mathematically to
make the numerical values of the scaled variables less stiff for the
numerical integration, and physically to see changes on the scale of
the age of our Universe. In order to achieve this goal, let us
introduce a dimensionless constant $T_0$, which represents the
magnitude of the Hubble time ($T_0=H^{-1}_0\doteq4.41504\times
10^{17}$ seconds). Then the scaling is
\begin{eqnarray}
\fl&&t\rightarrow T_0t,\quad a\rightarrow a,\quad H\rightarrow
H/T_0,\quad f\rightarrow f/T_0,\quad\chi\rightarrow\chi/T_0,\quad
R\rightarrow R/T_0^2,\quad E\rightarrow E/T_0^2,\nonumber\\
\fl&&A_0\rightarrow A_0,\quad A_2\rightarrow A_2,\quad
A_3\rightarrow A_3,\quad b^{+}\rightarrow T^2_0 b^{+},\quad
b^{-}\rightarrow T^2_0 b^{-},\quad
\Lambda\rightarrow\Lambda/T^2_0.\label{scale2}
\end{eqnarray}
With this scaling, all the field equations are kept unchanged while
the period $T\rightarrow T_0T$.
\subsection{The interaction between the scalar and pseudoscalar mode}
\label{inter2}
\begin{figure}[thbp]
\begin{tabular}{rl}
\includegraphics[width=0.47\textwidth]{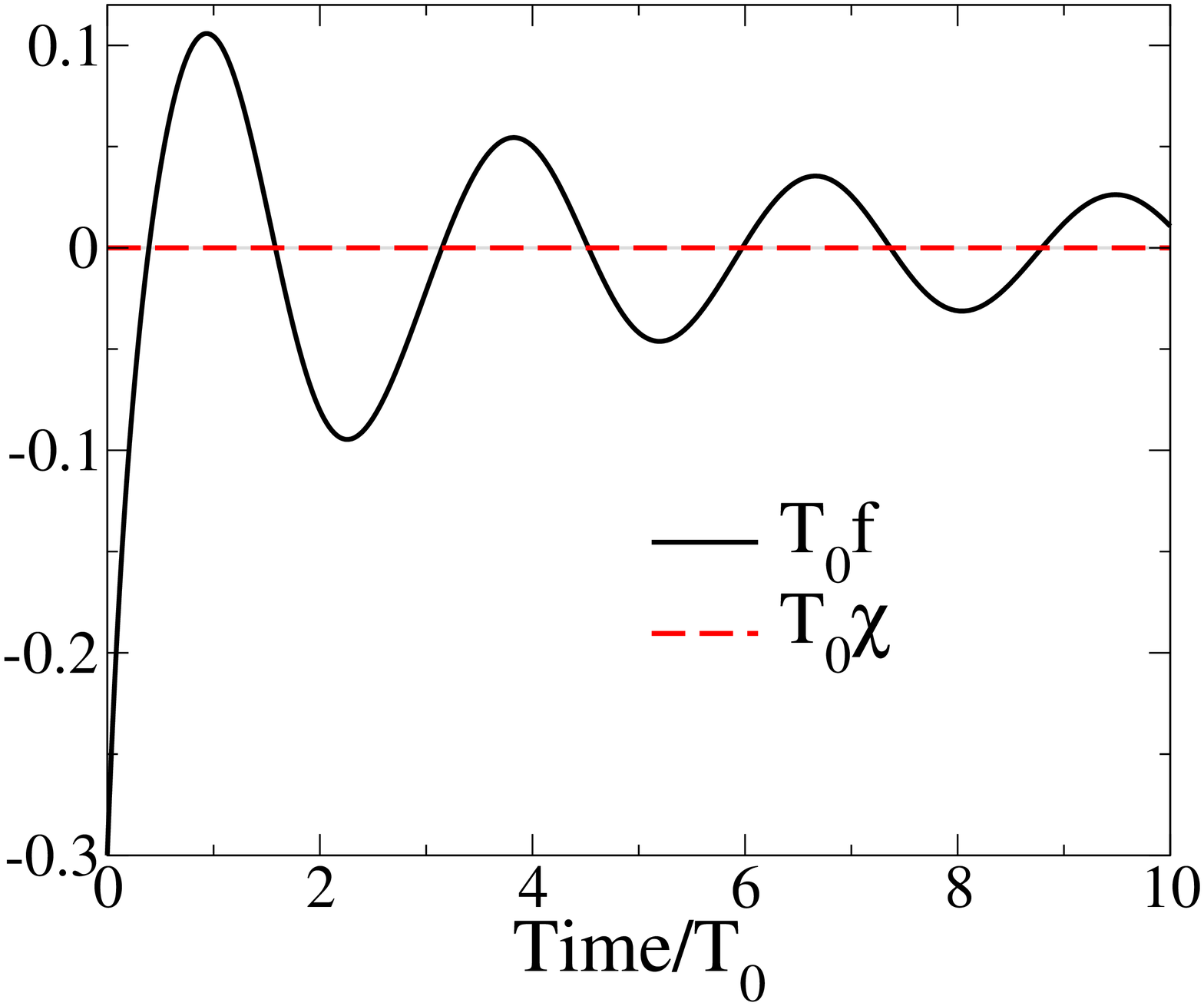}&
\includegraphics[width=0.47\textwidth]{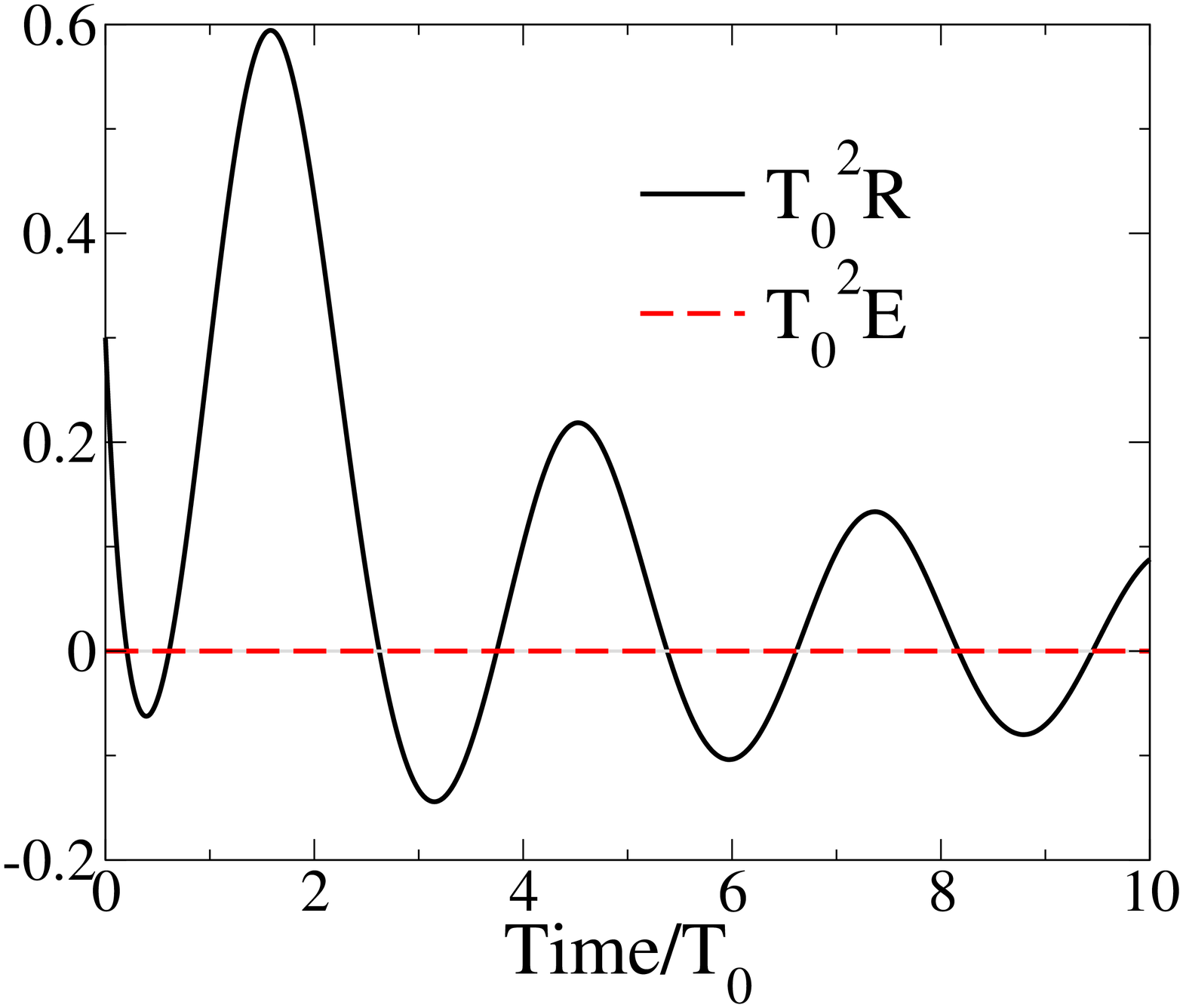}\\
\includegraphics[width=0.47\textwidth]{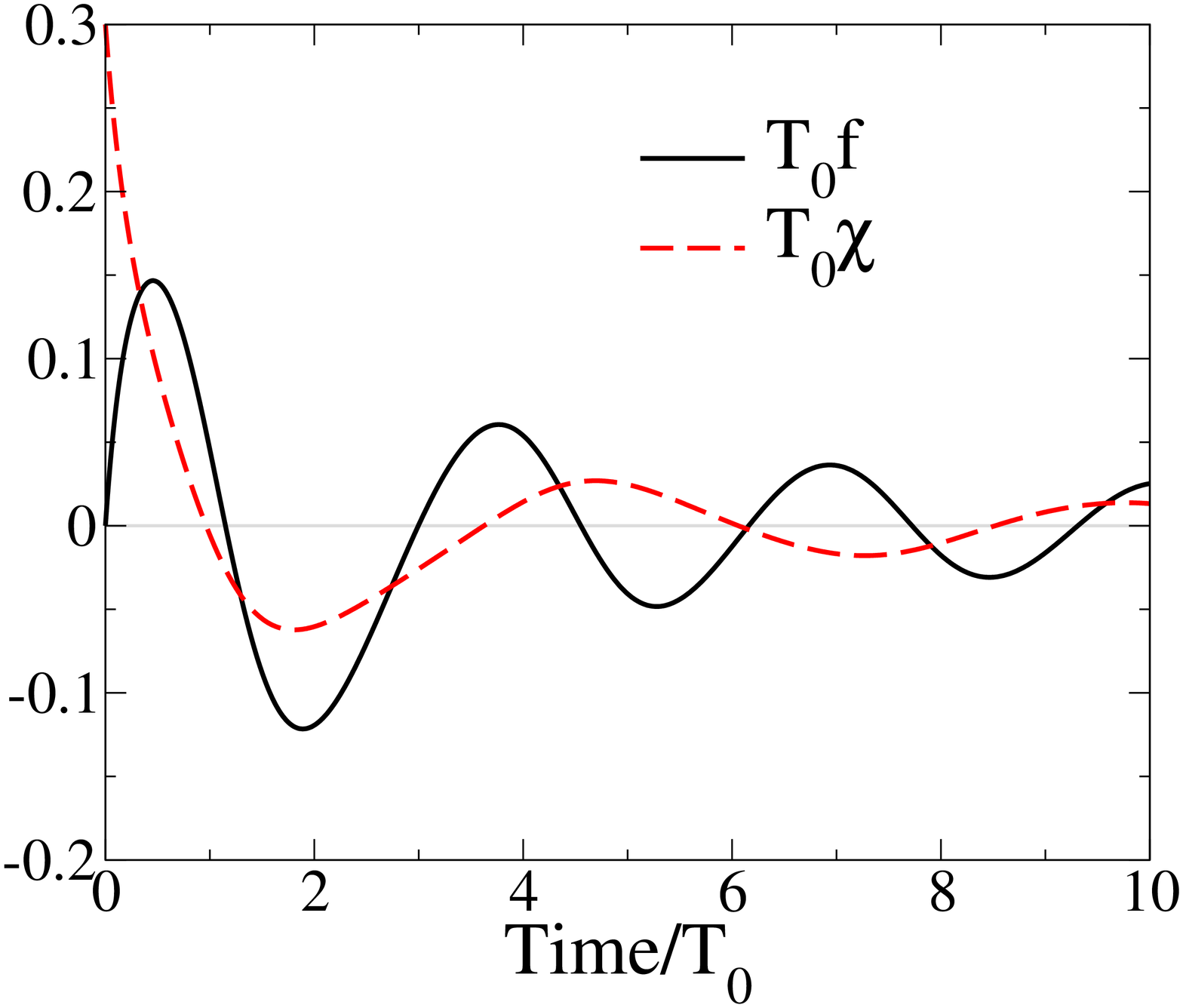}&
\includegraphics[width=0.47\textwidth]{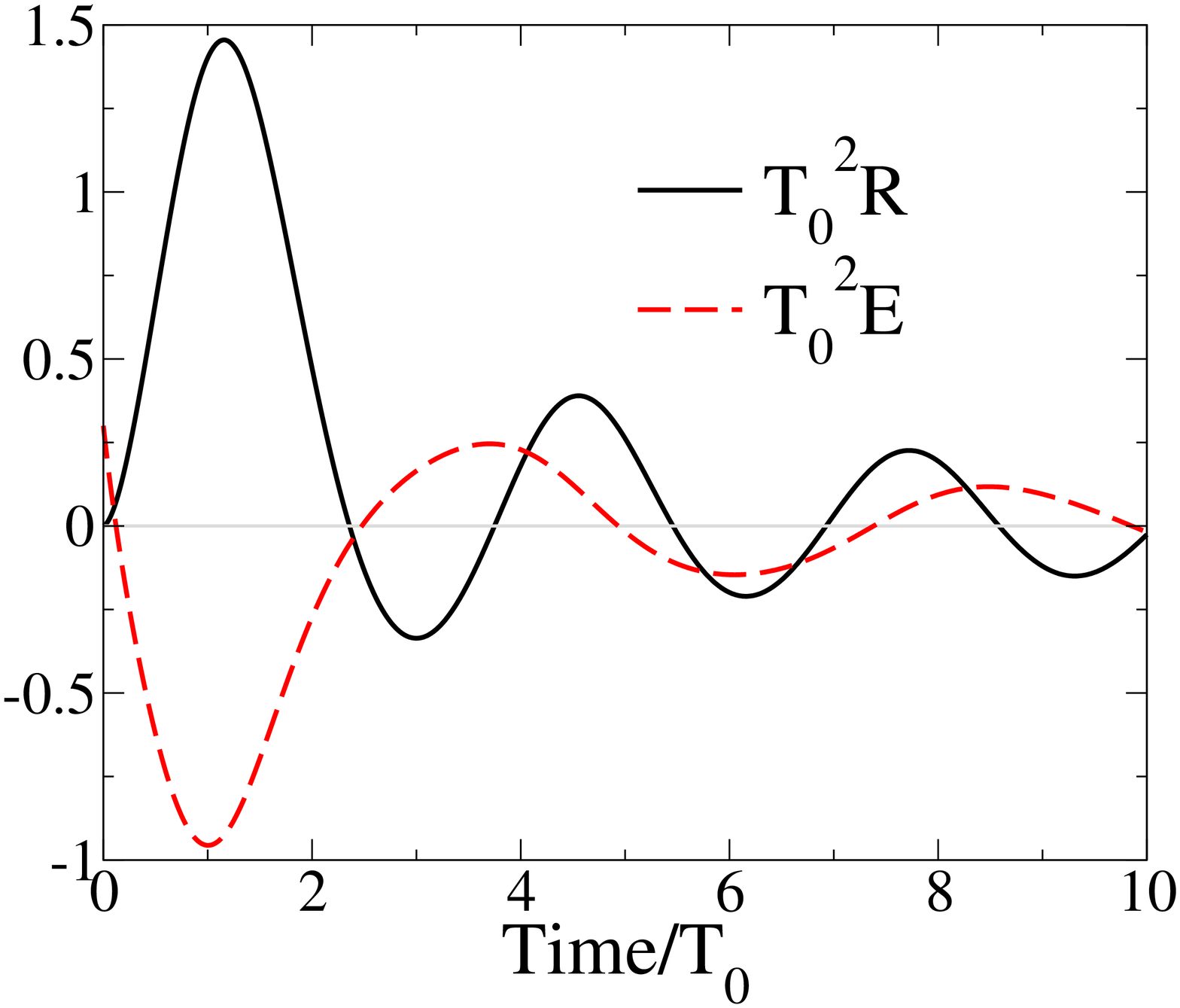}\\
\end{tabular}
\caption{Evolution of the components of the torsion. The top panels
show the evolution for nonvanishing $f$ and $R$, and vanishing
$\chi$ and $E$ initially, corresponding to Case (a). The bottom
panels show the evolution for nonvanishing $\chi$ and $E$, and
vanishing $f$ and $R$ initially, corresponding to Case (b).}
\label{ab}
\end{figure}
To understand the interaction of these two propagating scalar
connection modes, we consider two different situations: (a) the
pseudoscalar mode vanishes initially with a non-vanishing scalar
mode; (b) the scalar mode vanishes initially with a non-vanishing
pseudoscalar mode. The parameters and the initial values for Case
(a) are set as
\begin{equation}
A_2=0.5,\quad A_3=1,\quad b^{+}=2,\quad b^{-}=1,
\end{equation}
and
\begin{equation}
f(0)=-0.3,\quad R(0)=0.3,\quad\chi(0)=0,\quad E(0)=0,
\end{equation}
where $a(0)=50$ and $H(0)=1$ in both the cases. The parameters and
the initial values for Case (b) were chosen as
\begin{equation}
A_2=1.0,\quad A_3=-0.1,\quad b^{+}=1.5,\quad b^{-}=1,
\end{equation}
and
\begin{equation}
f(0)=0,\quad R(0)=0,\quad\chi(0)=0.3,\quad E(0)=0.3.
\end{equation}
The numerically calculated evolution of these two cases are shown in
Fig.~\ref{ab}. Case (a), which corresponds to the top two panels of
Fig.~\ref{ab}, shows that the scalar mode cannot generate an
initially vanishing pseudoscalar mode. It is clear that $\chi$ and
$E$ stay zero with a dynamic spin-$0^+$ mode. The two bottom panels
of Fig.~\ref{ab}, which correspond to case (b), show that the
pseudoscalar mode can generate an initially vanishing scalar mode.
It is known that the pseudoscalar connection mode will couple to
elementary spinning particles. One can expect that this mode will be
excited by spinning particles in the early Universe, since there
could be a high spin density during this epoch. Once the
pseudoscalar connection mode is generated, the scalar connection
mode will be excited through the interaction with the pseudoscalar
connection mode.
\subsection{Accelerating universes}
\label{preuni}
\begin{figure}[thbp]
\begin{tabular}{rl}
\includegraphics[width=0.47\textwidth]{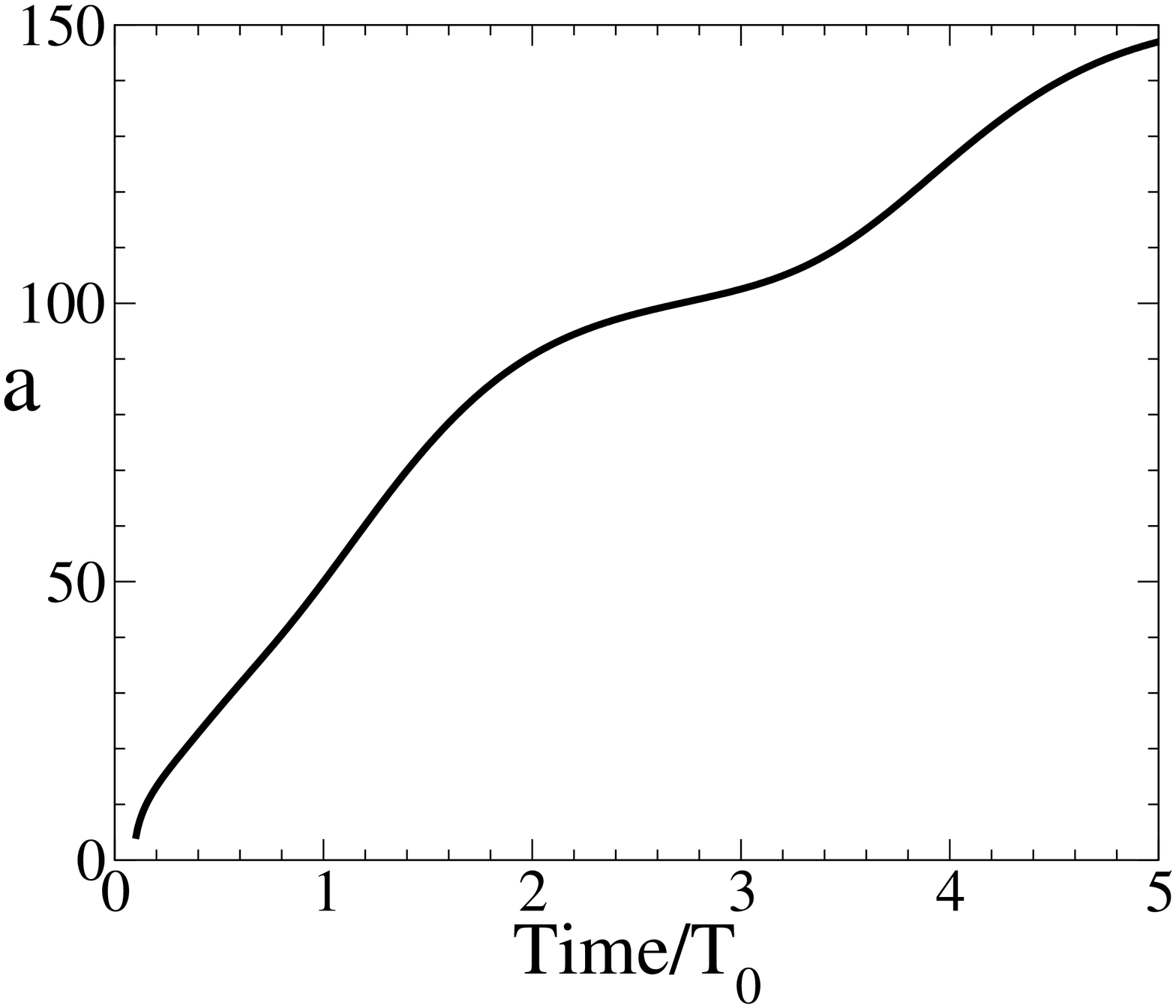}&
\includegraphics[width=0.47\textwidth]{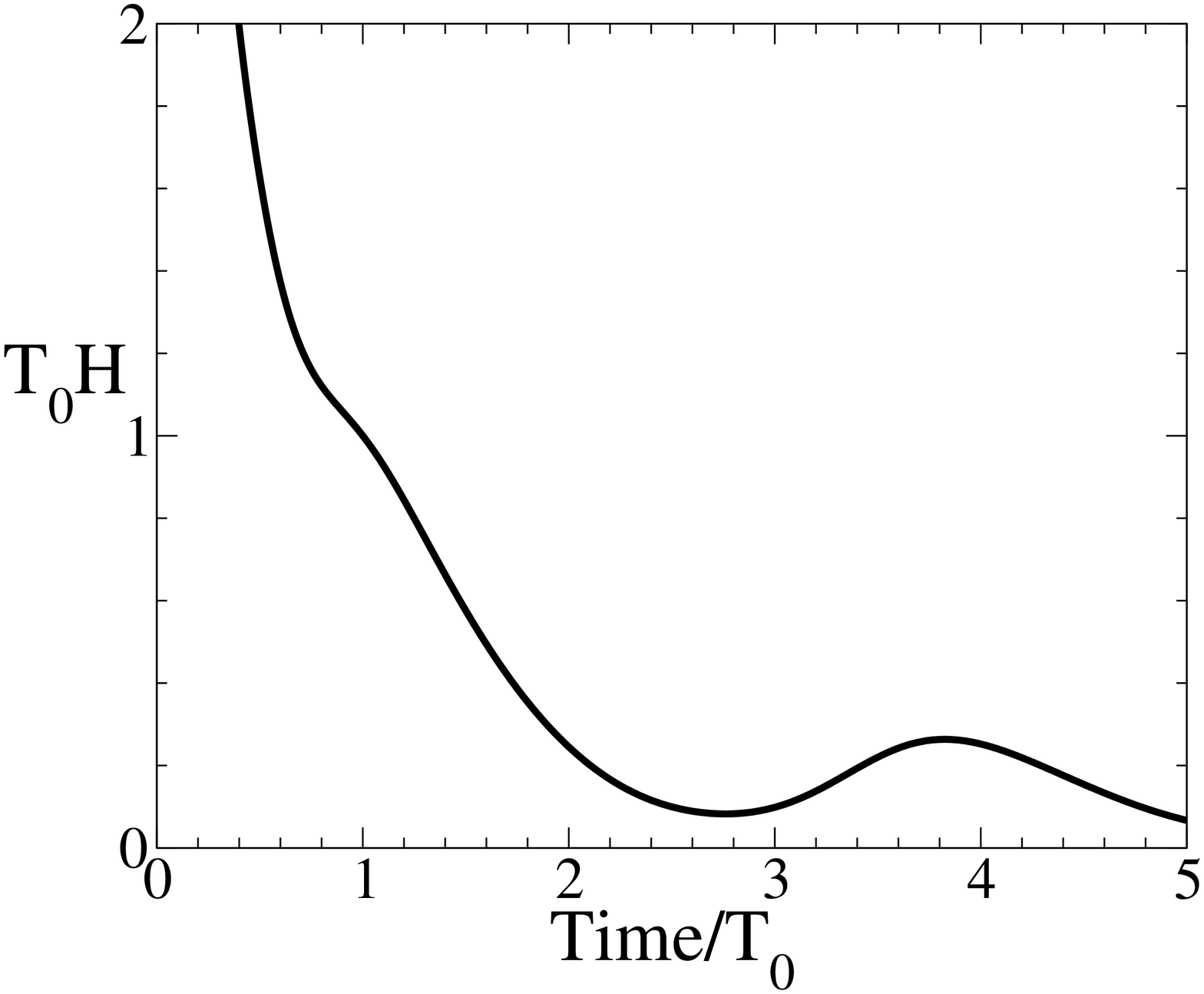}\\
\includegraphics[width=0.47\textwidth]{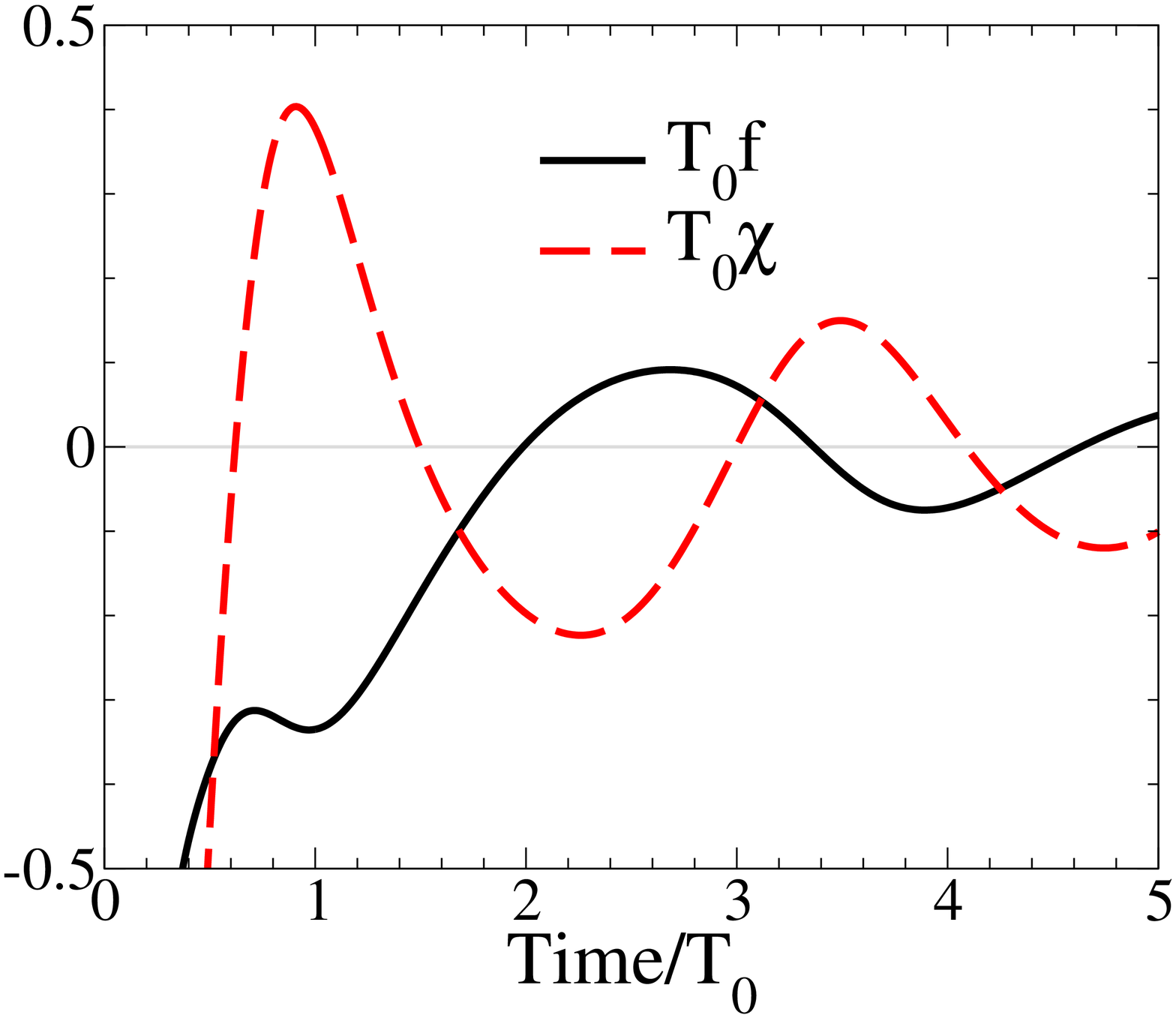}&
\includegraphics[width=0.47\textwidth]{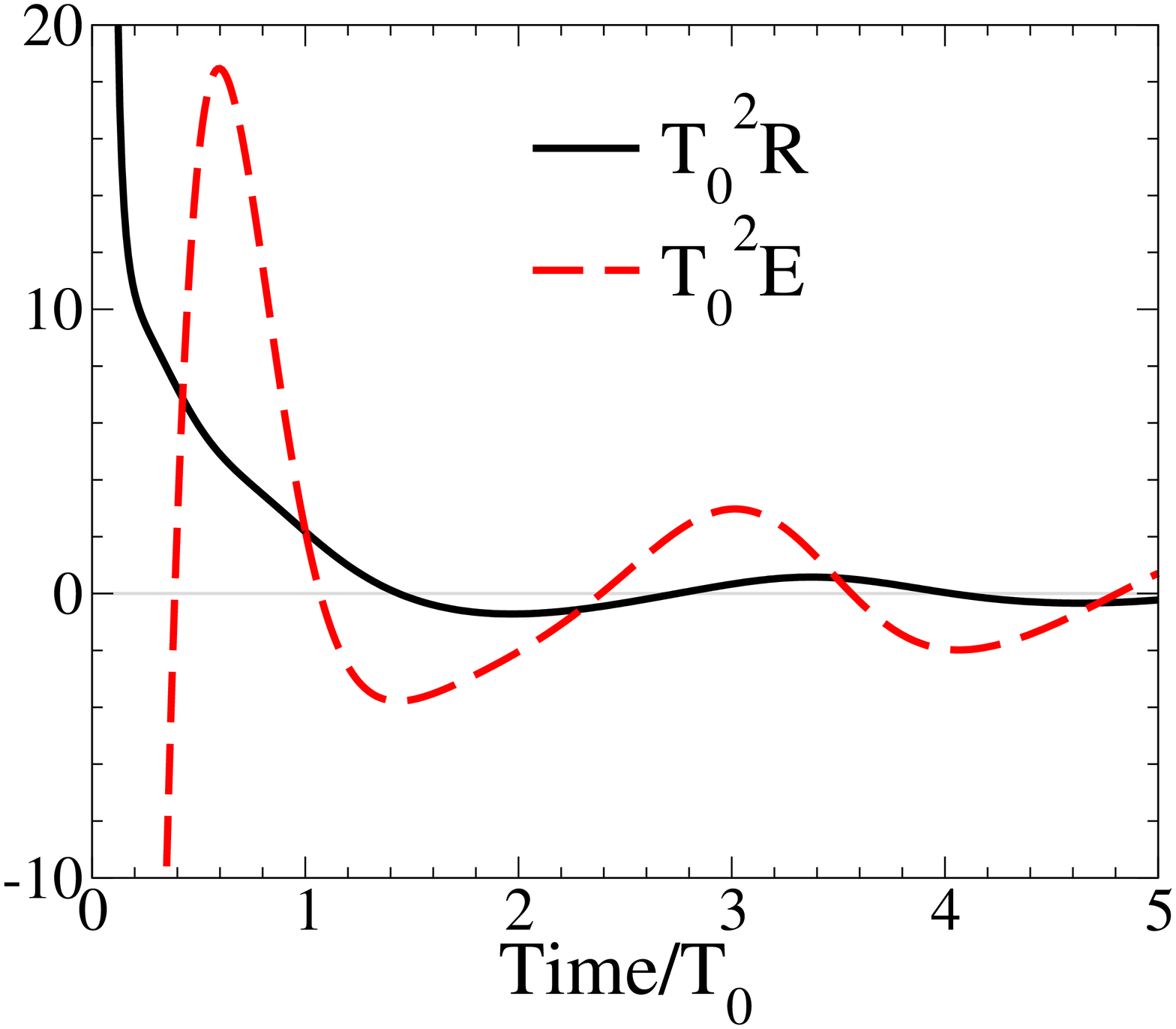}\\
\includegraphics[width=0.47\textwidth]{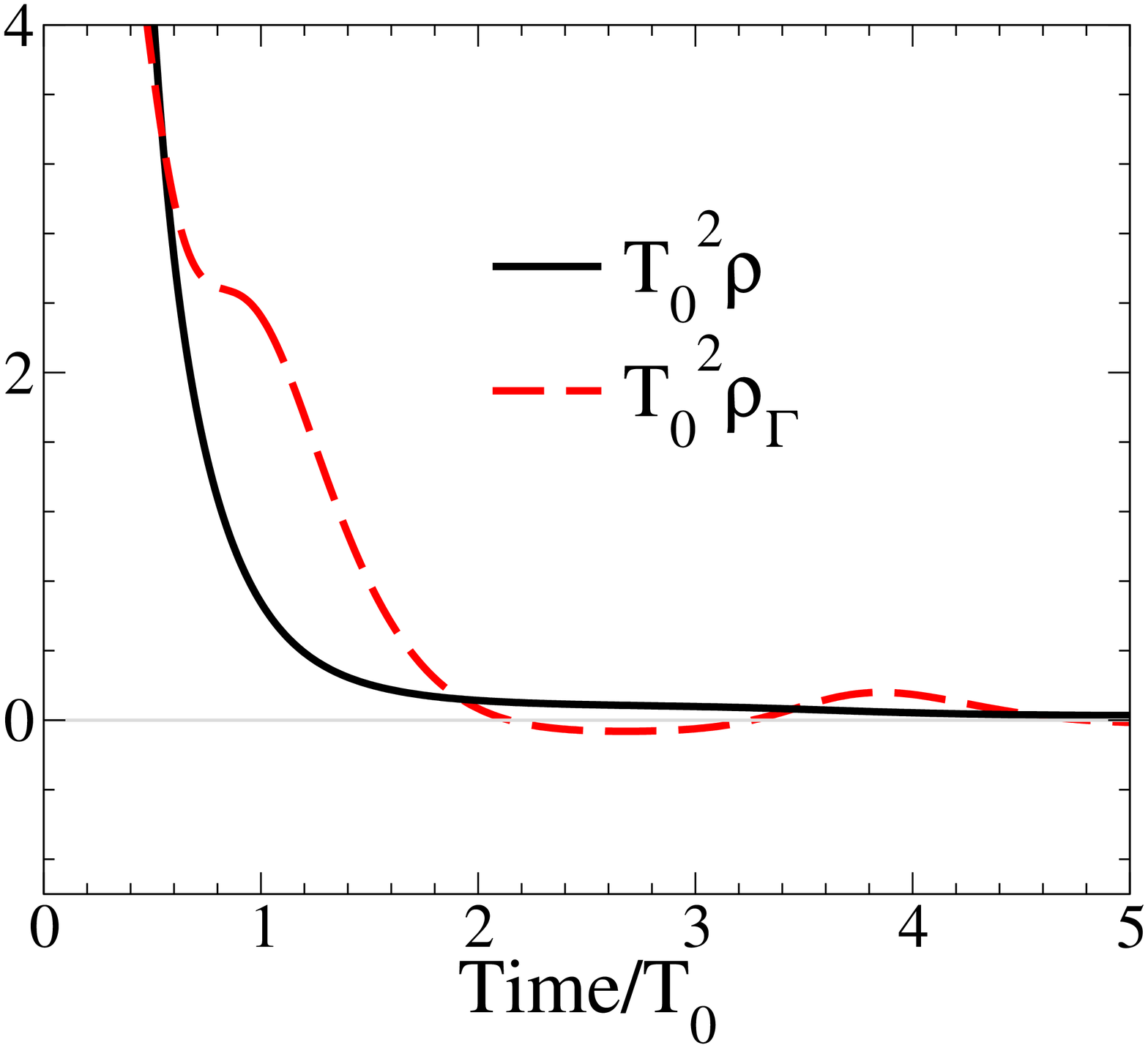}&
\includegraphics[width=0.47\textwidth]{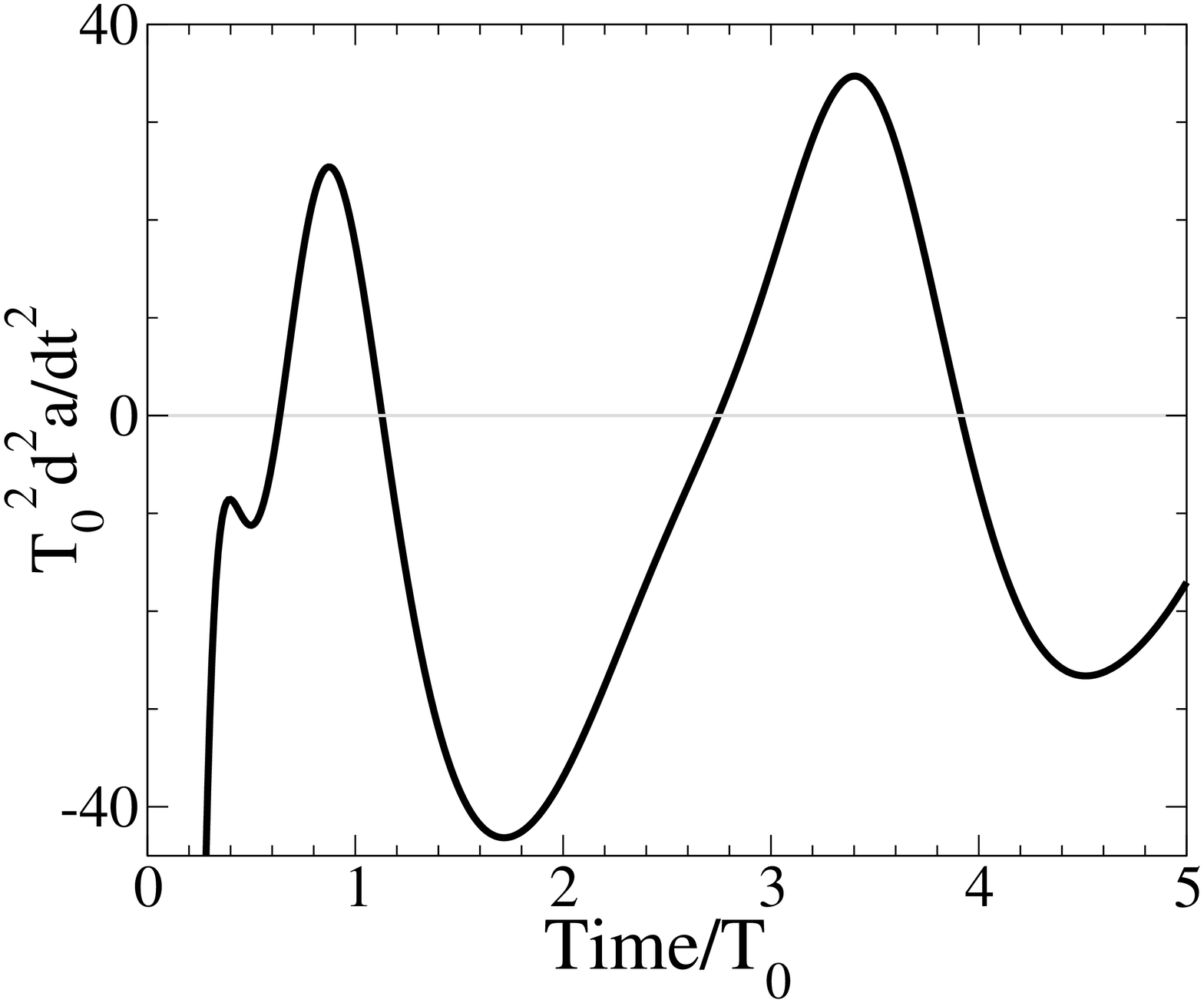}\\
\end{tabular}
\caption{Evolution of the expansion factor $a$, the Hubble function,
$H$, the scalar and the pseudoscalar torsion components, $f$ and
$\chi$, the affine scalar curvature, $R$, the pseudoscalar
curvature, $E$, the mass densities, $\rho$ and $\rho_\Gamma$, and
the 2nd time derivative of the expansion factor, $\ddot a$, as
functions of time with the parameter choice and the initial data in
Case I.} \label{fig3}
\end{figure}

\begin{figure}[thbp]
\begin{tabular}{cc}
\includegraphics[width=0.47\textwidth]{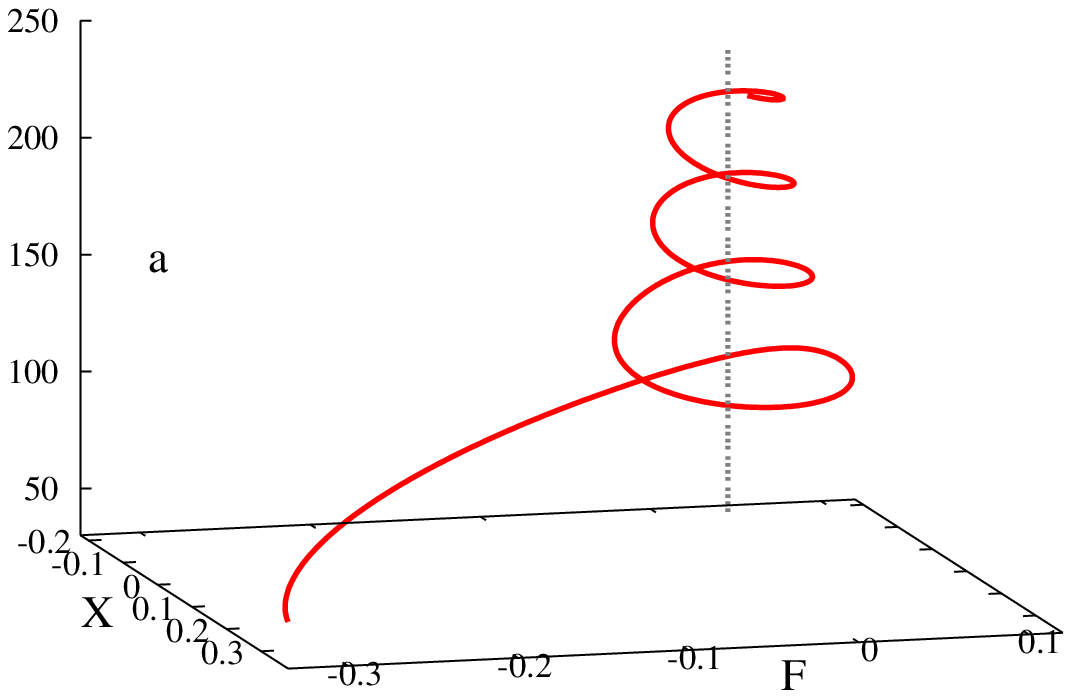}&
\includegraphics[width=0.47\textwidth]{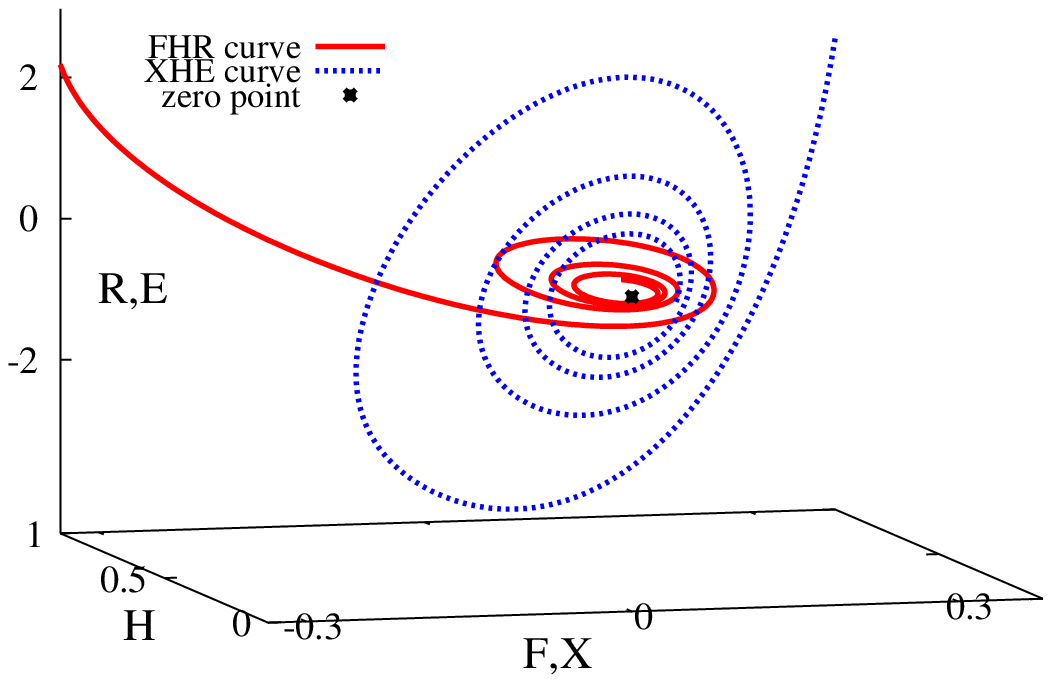}
\end{tabular}
\caption{The phase diagrams for Case I. The phase diagram of ($F$,
$\chi$, $a$) is shown in the left panel. The (red) solid line is the
trajectory of the ($F$, $\chi$, $a$) evolution starting from the
initial value ($-0.335$, $0.378$, $50$). The (gray) dashed line is
the convergence line ($0$, $0$, $a$) for this diagram. The phase
diagrams of ($F$, $H$, $R$) and of ($\chi$, $H$, $E$) are shown in
the right panel. The (red) solid line is the trajectory of the ($F$,
$H$, $R$) evolution starting from the initial value ($-0.335$, $1$,
$2.18$), the (blue) dashed line is the trajectory of the ($\chi$,
$H$, $E$) evolution starting from the initial value ($0.378$, $1$,
$2.21$), and the (black) filled point marks the asymptotic focus
point ($0$, $0$, $0$).} \label{fig4}
\end{figure}

We would like to compare the numerical evolution values for this
model (with $\Lambda=0$) with the observational data of our
Universe. The Hubble constant at present, $H(t_0=1)$, is
\begin{equation}
H=\frac{1}{4.41504\times 10^{17}}\cdot\frac{1}{\rm s}\approx 70
\frac{\rm km}{{\rm s}\cdot{\rm Mpc}}.
\end{equation}
The initial data is set at the current time $t_0=1$, and current
value of the Hubble function is scaled to unity in this work, just
as in \cite{SNY08}. The parameters and initial conditions chosen for
our first case are as follows:
\begin{equation}
A_2=0.83,\quad A_3=-0.35,\quad b^+=1.1,\quad  b^-= 0.091,
\end{equation}
and
\begin{eqnarray}
a(t_0=1)=50,\quad H(t_0=1)=1,\quad f(t_0=1)=-0.335,\nonumber\\
\chi(t_0=1)=0.378,\quad R(t_0=1)=2.18,\quad E(t_0=1) = 2.21.
\end{eqnarray}
The results of the evolution with these parameters and initial
conditions are plotted in Fig.~\ref{fig3}. The expansion factor $a$
is plotted in the top-left panel. In the top-right panel the Hubble
function $H$ is damped-oscillating at late time. In the bottom-right
panel, it is obvious that $\ddot{a}$ is damped and oscillating
during the evolution and is positive at the current time $t\approx
1$, which means the expansion of the universe is currently
accelerating. The torsion and curvature scalars, $f(t)$, $\chi(t)$,
$R(t)$, and $E(t)$, are also plotted in the middle panels of
Fig.~\ref{fig3} to show the correlation of the evolution between
these variables. We observe that the frequencies of the pairs
($\chi$,$E$) and ($f$,$R$) are usually different. The behavior is
consistent with the analysis in Sec.~\ref{asympexp}. In order to
have a deeper understanding of the settings of this case, the matter
density $\rho$ and the effective mass density of the dynamical
connection $\rho_\Gamma$ are plotted in the bottom-left panel. The
value of $\rho$ is decreasing as the universe is expanding and is
always positive with $\rho a^3=$const,
 while $\rho_\Gamma$, plotted in the same panel,
shows a ``damped-oscillating'' behavior. The damped-oscillating
behavior of $\rho_\Gamma$ simply indicates that the effective energy
density $\rho_\Gamma$ is not positive-definite in general. We also
plot the phase diagrams for Case I in Fig.~\ref{fig4} to show that
the orbit of ($f$,$\chi$,$a$) is convergent to a line ($0$,$0$,$a$),
and the orbits of ($f$,$H$,$R$) and of ($\chi$,$H$,$E$) both
converge to the point ($0$,$0$,$0$).

\begin{table}[t]
\caption{The initial data and parameters for cases I, II, and III.
Here the parameter $A_0=1$, $a(t=1)=50$, and $H(t=1)=1$ in all three
cases; $t=1$ means $t={\rm now}$, under the scaling of
Eqs.~(\ref{scale1}--\ref{scale2}).}
\begin{tabular}{cccccccccc}
\hline Case
&$A_2$&$A_3$&$b^{+}$&$b^{-}$&$f(1)$&$\chi(1)$&$R(1)$&$E(1)$&
$\Omega_m$\\
\hline
I  & 0.83  & -0.35 & 1.1  & 0.091& -0.335 & 0.378 & 2.18 & 2.21 &0.23\\
II & 0.52  & 0.475 & 1.05 & 0.35 & -0.318 & 0.225 & 2.7  & -1.2 &0.29\\
III& 0.635 & 0.5   & 1.06 & 0.42 & -0.361 & 0.058 & 2.442& -1.8 &0.27\\
\hline
\end{tabular}
\label{allpar}
\end{table}

\begin{figure}[t]
\begin{tabular}{cc}
\includegraphics[width=0.94\textwidth]{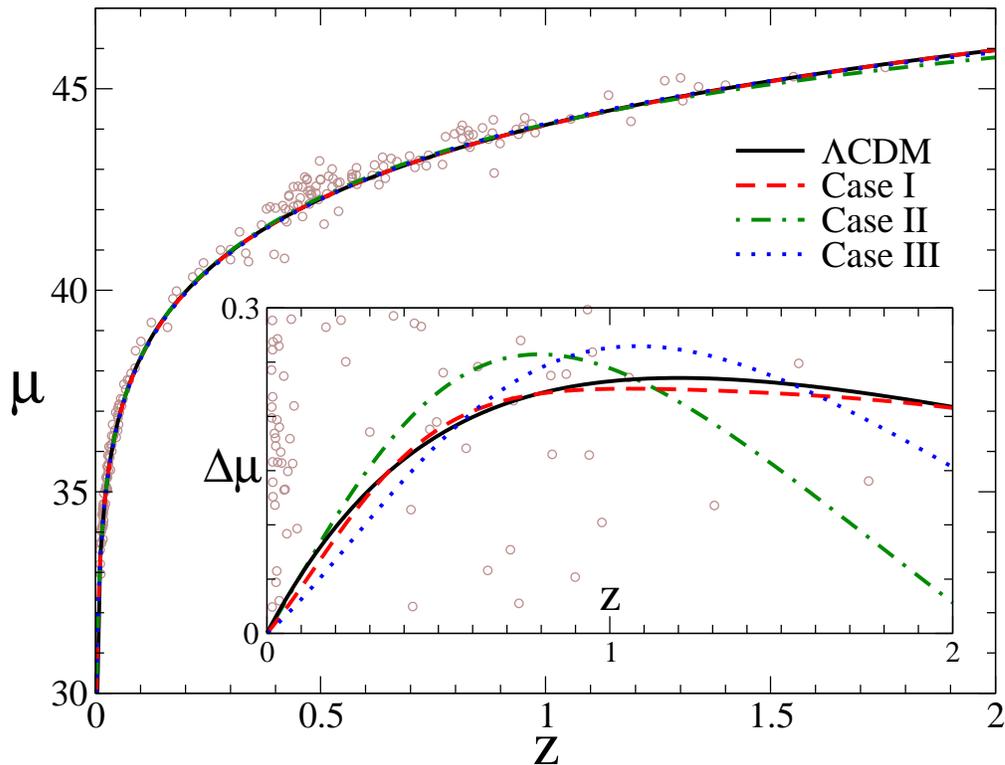}
\end{tabular}
\caption{Comparison of different spin-zero connection models and the
standard $\Lambda$CDM model with the observational data via the
relation between the distance modulus $\mu$ and the redshift $z$.
The supernovae data points, plotted with (brown) circles, come from
\cite{Reis04}. The result of the standard $\Lambda$CDM model
($\Omega_{\rm m}=0.3$, $\Omega_\Lambda=0.7$) is plotted by the bold
solid line.  The results of Case I, II, and III are represented by
the (red) dashed line, the (green) dot-dashed line, and the (blue)
dotted line, respectively. In the inset, the models and data are
shown relative to an empty universe model ($\Omega=0$). }
\label{fig5}
\end{figure}

In this case the scaled value of $\rho(t=1)=0.68$ and its physical
value is $\rho(t=T_0)=2.15\times 10^{-30}{\rm g}/{\rm cm}^3$. The
Universe is supposed to be very close to the critical density,
$\rho_c\equiv3c^2H^2/8\pi G=9.47\times 10^{-30}{\rm g}/{\rm cm}^3$;
we find the ratio $\Omega_{\rm m}\equiv\rho/\rho_c\approx 23\%$. In
the standard $\Lambda$CDM model, $\Omega_{\rm m}\sim30\%$ with $5\%$
baryonic matter and $25\%$ dark matter. For our model $\Omega_\Gamma
\equiv\rho_\Gamma/\rho_c=77\%$ acts like the energy density of dark
energy. Therefore, this dynamic connection model is able to describe
a presently accelerating expansion of the Universe with a proper
amount of matter density. From the field equations  we can see that
the {\it effect} of the ``dark energy" mainly comes from the
nonlinearity of the field equation driven by the dynamic scalar and
pseudoscalar connection modes. We also found other cases, two of
which are listed in Table~\ref{allpar} along with Case I; they are
obtained by taking different values for the parameters and the
initial conditions. We find that the results for the other two cases
have a behavior qualitatively similar to that of Case I.

We compare our results with the supernovae data. Distance estimates
from SN Ia light curves are derived from the luminosity distance
\begin{equation}
d_L\equiv\sqrt{\frac{L_{\rm int}}{4\pi{\cal F}}}=cT_0 a(1)(1+z)
\int^t_1\frac{{\rm d}t}{a(t)}\,,
\end{equation}
where $L_{\rm int}$ and $\cal F$ are the intrinsic luminosity and
observed flux of the SN, and the redshift $z\equiv a(1)/a(t)-1$.
Logarithmic measures of the flux (apparent magnitude, $m$) and
luminosity (absolute magnitude, $M$) were used to derive the
predicted distance modulus
\begin{equation}
\mu=m-M=5\log_{10}d_L+25\,,
\end{equation}
where $m$ is the flux (apparent magnitude), $M$ is the luminosity
(absolute magnitude), and $d_L$ in the formula should be in units of
megaparsecs. We found the relations between the predicted distance
modulus $\mu$ and the redshift $z$ in the three cases; they are
plotted in Fig.~\ref{fig5}. For comparison, we also plot the
prediction of the $\Lambda$CDM model with $\Omega_{\rm m}=0.3$ and
$\Omega_{\Lambda}=0.7$ by employing the following formula
\cite{Reis04}
\begin{equation}
d_L=cT_0(1+z)\int_0^z\frac{{\rm d}z}{\sqrt{(1+z)^2(1+ \Omega_{\rm
m}z) -z(2+z)\Omega_\Lambda}}\,.
\end{equation}
The astronomical observational data \cite{Reis04,SNIaRP} are also
plotted in Fig.~\ref{fig5} for comparison. The plots show that for
small redshift $z$ (e.g., $z<1.9$) all three cases of the dynamical
connection models give an accelerating universe just like the
$\Lambda$CDM model does. For larger $z$ these cases might turn the
Universe into a deceleration mode, which is consistent with the
behavior of the various quantities shown in Fig.~\ref{fig3}. We can
see that Case I gives the closest curve behavior to the one from the
$\Lambda$CDM model. In Fig.~\ref{fig5}, we demonstrate the
possibility of the spin-zero connection fields accounting for the
effect of dark energy with a suitable set of parameters and initial
data. A comparison of Fig.~\ref{fig5} with the results in
\cite{SNY08} shows that this two-scalar-mode model can give (not
surprisingly) a better fitting of the supernova data than the
one-scalar-mode model can.

\section{Discussion}
From a series of earlier works
\cite{HNZ96,CNY98,yo-nester-99,yo-nester-02} it was concluded that
the Poincar\'e Gauge Theory of gravity has two good dynamic Lorentz
connection modes, the ``scalar'' mode (spin $0^+$) and the
``pseudoscalar'' mode (spin $0^-$) which satisfy 2nd order
equations.

Here we extended a previous work \cite{SNY08}, which considered a
PGT cosmological model with one dynamic Lorentz connection mode
having spin $0^+$, to the case where both the scalar and
pseudoscalar connection modes are dynamic. The objectives are (i) to
study this PGT cosmological model (and in particular how well it can
match the present universe observations) and (ii) to get a deeper
understanding of the dynamics of the PGT.

From the cosmological homogeneous and isotropic assumptions the
scalar and pseudoscalar curvatures $R,E$ and the temporal components
of the trace torsion $f$ and axial torsion $\chi$ survive and affect
the evolution of the universe in this two-connection-mode model.
Recognizing the equivalence of the model to one describing a
particle with three degrees of freedom, we constructed an effective
Lagrangian and the corresponding Hamiltonian by imposing the FLRW
symmetry on the field theory action. The system of ODEs obtained
therefrom are the same as the evolution equations obtained by
imposing the FLRW symmetry on the equations derived from the PGT
Lagrangian density.

With the evolution equations (\ref{dta})--(\ref{dtE}) and the
associated energy constraint (\ref{constraint}) we analyzed the late
time asymptotic expansion. We found there are three normal modes:
one related to the Hubble expansion and two dynamic modes
represented by the torsion and curvature components. It was found
that only the scalar mode affects the late-time expansion rate. The
numerical analysis focused on the interaction between these two
modes, the study of the possible behavior of the Universe, and the
fitting to the observed supernova data. It was shown that the
dynamical activity of the pseudoscalar mode could excite the scalar
mode via the nonlinear coupling of these two modes, but the converse
does not happen: one can have the scalar mode excited without any
pseudoscalar excitation. Like the one-mode model in \cite{SNY08},
the present model allows for an expanding universe with an
oscillating component in the expansion rate.  Consequently, although
on the average the expansion is slowing down, the universe can have
an accelerating expansion at the present time. The additional degree
of freedom in this two-mode model compared to the one in
\cite{SNY08}, not surprisingly, allows us to obtain a better fit to
the supernova data.

From the evolution equations~(\ref{dtf})--(\ref{dtE}), we can see
the nonlinear coupling between the scalar and pseudoscalar modes. In
Sec.~\ref{inter2} we demonstrated the excitation of the scalar mode
through the dynamical activity of the pseudoscalar mode. Such a
nonlinear interaction between these two modes offers a natural
mechanism to fuel the strength of the scalar mode in the evolution
of the Universe. We stress that there is no known fundamental
material source which directly excites the scalar mode; this part of
the Lorentz connection simply does not interact in any obvious
fashion
 with any familiar type of matter \cite{ShaI02}.
Conversely, the pseudoscalar mode is naturally driven by the
intrinsic spin of fundamental fermions; in turn it naturally
interacts with such sources. Indirectly, the $0^+$ mode could be
enhanced and activated dynamically through the aforementioned
non-linear mechanism, in addition to any possible primordial
amplitude from the early universe.

From the late-time analysis in Sec.~\ref{asympexp}, we showed that
only the scalar mode, not the pseudoscalar mode, plays a direct role
in affecting the expansion rate of the Universe. This result is
perfectly consistent with our understanding of the characteristics
of these two modes: Due to the ability of interacting with fermionic
matter, it is generally thought that the axial torsion (controlled
by the pseudoscalar part of the Lorentz connection) must be small
and have small effects at the present time \cite{CSFG94}.
Conversely, the scalar Lorentz connection mode could be considered
as a ``phantom" field, at least in the matter-dominated epoch, since
it will not interact directly with matter, and yet can drive the
Universe in an oscillating fashion with an accelerating expansion at
the present time.

As discussed in \cite{SNY08}, the two Lorentz connection spin-zero
modes in this model are in some ways effectively like a scalar field
and a pseudoscalar field, yet these two ``scalar'' fields are
fundamentally different from the various scalar field models of
unknown matter, e.g., the quintessence models, in the following
ways:
\begin{itemize}
\item this cosmological model is derived naturally from a geometric
gravitational theory, the PGT, which is based on fundamental gauge principles,
instead of on the hypothesis of the existence of a dark  energy
tailored to producing an explanation of an accelerating universe;
\item  there are, consequently, only a few free parameters in this cosmological
model, instead of an ad hoc potential that can be rather arbitrarily chosen
to fit the observations. Therefore, this PGT cosmological model
should be more restrictive, and should be easier to be confirmed or
falsified;
\item based on its geometric character, the
coupling of the dynamic parts of the Lorentz connection to the other
fields is nothing like that which has ever been advocated for
hypothetical scalar fields.
\end{itemize}
Thus this PGT cosmology with a Lorentz connection having dynamic
``scalar'' modes and the quintessence models are characteristically
different, even though there are some similarities.

As mentioned in our previous work, if we consider the spacetimes as
Riemannian instead of Riemann-Cartan, by absorbing the contribution
of the post-Riemannian terms of this model into the stress-energy
tensor on the rhs of the Einstein equations, as indicated in
(\ref{energy_effective},\ref{rhoGamma},\ref{pressure_effective},\ref{pGamma}),
then this contribution will act as a source of the Riemannian
metric, effectively like an {\em exotic} fluid with its mass density
$\rho_{\Gamma}$ and pressure $p_{\Gamma}$ varying with time
(although the time evolution of these torsion and curvature terms
are not like that of any fluid). Moreover, the effective fluid
will appear to have presently a negative pressure, and consequently
a negative parameter in the effective equation of state, i.e.,
$\omega_{\Gamma}\equiv p_{\Gamma}/\rho_{\Gamma}$, which drives the
universe into accelerating expansion. Note that there is no
constraint on the value of $\omega_{\Gamma}$ which appears here, and
its value could vary from time to time. It should be stressed that
this is not a real physical fluid situation; the truth is that
$\omega_{\Gamma}$ is nothing like ``a connection field equation of
state'', it is just a proportionality factor between $\rho_{\Gamma}$
and $p_{\Gamma}$, two expressions which effectively summarize the
contribution of the connection (via the curvature and torsion)
acting as a source of the metric. The ratio $\omega_{\Gamma}$ is of
interest only to help understand the acceleration of this model and
to enable a limited comparison with other dark energy proposals.

By imposing the FLRW symmetry on the Lagrangian density, we
constructed an effective Lagrangian as well as the corresponding
Hamiltonian. One benefit of the former is a simpler derivation of
the dynamic equations (\ref{dta})--(\ref{dtE}).  The latter should
also prove useful,  as the Hamiltonian formulation is the framework
for the most powerful known techniques for analytically
investigating the dynamics of a system. By these techniques one can
to apply the experience accumulated in dealing with conservative
classical mechanical systems.  An effective Lagrangian and the
corresponding Hamiltonian allows one to visualize the system as a
particle moving in a potential.  This would be very helpful in
gaining a better appreciation of the dynamics of any sophisticated
model.
 (Note, it is not necessarily true that an
effective Lagrangian can be found in an arbitrary cosmological
model. Extrapolating from  GR, one can conjecture that this is
possible for all PGT Class A  Bianchi models with suitable sources,
including pressure and spin). As we have seen in
Sec.~\ref{asympexp}, the effective mechanical system methods were
also useful for the late time normal mode analysis.

There have been some studies on PGT cosmology with dynamic scalar
connection modes since the model proposed in \cite{YN07,SNY08}. Wang
and Wu \cite{WW09} considered a related, but fundamentally different
model, which turns out to have only the dynamic $0^-$ mode.  They
considered the early universe and showed how in their model such a
dynamic PGT connection could account for inflation (for another
approach to using the PGT to account for inflation, see
\cite{MG06}). Li {\it et al.}~\cite{LSX09,LSX09b} presented a nice
analysis of the scalar mode model of \cite{SNY08} from a more
mathematical angle in order to get a deeper insight into the
behavior of the dynamical system. In their work, they found the
critical points of the system and the corresponding ranges of the
parameters. In the latter work they also fit the model to the
supernova data to find the best fit values of the parameters.
These works considered quite general ranges of the parameters and
found several interesting dynamical effects. We note that many of
these interesting effects happen in parameter ranges which are
outside of the restrictions considered by \cite{SNY08} to be
physically necessary in order to have good linear modes (long ago
\cite{HS80,SN80} the conditions were found so that the propagating
modes should carry positive energy and satisfy the
no-faster-than-light condition). Also, a matter density of about
$25\%$ of the critical density was also imposed in \cite{SNY08} to
give a more physical meaning to the a curve fitting. Further
investigations and a careful study of the model will be needed to
pin down the acceptable ranges of the parameters. In the present
work, we have chosen the range of the related parameters following
the result of \cite{yo-nester-99} for good propagating scalar linear
modes, just as in \cite{SNY08}. Applying the methods used in
\cite{LSX09,LSX09b} to the present extended model would surely lead
to further insights.

One may wonder: how large must the post-Riemannian fields be in
order to produce observable effects in the the present day universe,
e.g., the observed acceleration? Conversely, how large can the
torsion or curvature scalars be without violating some observational
constraint? The questions merit a detailed study.  Here is a simple
argument that indicates a magnitude. Let us compare the terms in the
Lagrangian density and the field equations for the model in which
the PGT Lorentz connection has scalar dynamical modes and the
Einstein theory with a cosmological constant. (In our present work
we have deliberately included in most of the dynamical equations a
possible cosmological constant; this was done not only for greater
generality but also to facilitate just such a comparison.  In our
numerical evolution for our model we used $\Lambda=0$.) Note that
the presumed cosmological constant is ``so small'' that it has no
noticeable effect in the laboratory, nor on the solar system scale,
nor on the galactic scale. Nevertheless it is large enough to have
the dominant effect on the cosmological scale. Hence we are led to
infer that we should consider that one or more of the
post-Riemannian terms ($A_2f^2$, $A_3\chi^2$, $b^+R^2$, $b^-E^2$)
should be comparable to the cosmological constant (which is about
$3\rho$) in the $\Lambda$CDM model. With such a choice we can expect
that the post-Riemannian terms may be able to accelerate the
universe and yet not be conspicuous on smaller scales.

The introduction of a new ingredient (i.e., the $0^-$ connection
mode which is reflected in the axial torsion and the pseudoscalar
curvature) in this work raises the concern of the experimental and
observational constraints on this field.
There have also been some laboratory tests in search of torsion
\cite{CTNW93,NiWT96}. The main idea among these experiments is the
spin interaction between matter and torsion. The theoretical
analyses and the high energy experimental data on four-fermion
vertices sets the lower bound for the (pseudoscalar) torsion mass
$>200$ Gev \cite{CSFG94,BSSI99,ShaI02,CLSC03}.
The cosmological
tests on torsion have investigated the effect of torsion-induced
spin flips of neutrinos in the early Universe---which could alter
the helium abundance and have other effects on the early
nucleosynthesis \cite{CILS99,BruM99}. From Table \ref{allpar},
the parameters chosen for the range of the the torsion
mass are consistent with the aforementioned analyses.
Our model is also comfortable with the most restrictive
experimental limits found on torsion \cite{KRT08}.
For torsion being applied to the cosmological problem, Capozziello {\it et
al.}~\cite{CaCT03,CSet03} have done a serious study on replacing the
role of the cosmological constant in the accelerating Universe. With
a totally antisymmetric torsion without dynamical evolution, their
model is consistent with the observational data by tuning the amount
of the torsion density. Compared with them, the model in this work
allows the pseudoscalar torsion to evolve dynamically. This
difference might enable a more fruitful physics to be studied.


\section{Conclusion}
In this work we considered the two ``scalar'' dynamical modes of the
PGT Lorentz connection in a cosmological setting and have proposed
it as a viable model for explaining the current status of the
Universe. Besides seeking a better understanding of the PGT, we have
considered the prospects of accounting for the outstanding present
day mystery---the accelerating universe---in terms of an alternative
gravity theory, more particularly in terms of the PGT with a dynamic
Lorentz connection having only two dynamic modes, carrying spin-0
with even and odd parity. With the usual assumptions of isotropy and
homogeneity in cosmology, we find that, under the model, the
Universe will have with generic choices of the parameters an
expansion rate which oscillates. The connection in this model could
play the role of dark energy. With a certain range of parameter
choices, it can account for the current status of the Universe,
i.e., an accelerating expanding universe with a value of the Hubble
constant which is approximately the present one. Thus we have
considered the possibility that a certain geometric field, a dynamic
Lorentz connection---which is naturally expected from spacetime
gauge theory---could fully account for the accelerated universe.

The $0^+$ mode, which directly drives the acceleration of the
universe, does not couple directly to any known material source. By
way of non-linear terms it could come indirectly from the huge
density of the particles with sufficient spin alignment in the early
universe which directly excite the $0^-$ connection mode. The $0^+$
mode could be considered as a ``phantom" field, at least in the
matter-dominated epoch, since it will not interact directly with
matter; it only interacts indirectly via the gravitational
equations. Then the dynamics of the scalar torsion mode could drive
the Universe in an oscillating fashion with an accelerating
expansion at present. It is quite remarkable that a gauge theory of
dynamic geometry naturally presents us with such a ``phantom''
field.  This natural geometric field could act like a dark energy.

\section*{Acknowledgments}
This work was supported in part by the National Science Council of
the R.O.C. (Taiwan) under grant Nos.~NSC97-2112-M-006-008,
NSC97-2112-M-008-001. This work was also supported in part by the
National Center of Theoretical Sciences and the (NCU) Center for
Mathematics and Theoretical Physics. Some of the calculations were
performed at the National Center for High-performance Computing in
Taiwan.  The encouragement and helpful advice of F. W. Hehl and C.
Soo was much appreciated.
\section*{Appendix A: The choice of parameters}
Regarding our choice of parameters.  From the table which can be
found in any one of \cite{CNY98,yo-nester-99,yo-nester-02}, we find
that to kill the dynamics of the $1^+$, $2^+$, $1^-$, $2^-$ modes we
want to take, respectively,
\begin{equation}
 b_2+b_5=0,\quad b_1+b_4=0,\quad b_4+b_5=0,\quad b_1+b_2=0.
\end{equation}
For dynamic $0^+$ and $0^-$ we want, respectively,
\begin{equation}
b_4+b_6>0,\quad{\rm and}\quad b_2+b_3<0,
\end{equation}
from Table 3 in \cite{yo-nester-02}.
Now, due to the Bach-Lanczos identity, we can
choose any one of the parameters $b_k$ to vanish. Taking, say
$b_4=0$ we then get that we also want $b_1=b_2=b_5=0$, leaving
$b_6>0$, $b_3<0$.
We also find for the dynamic $0^+$ and $0^-$ modes,
from Table 3 in \cite{yo-nester-02} the respective restrictions
\begin{equation}
a_0a_2(2a_0+a_2)<0,\quad{\rm and}\quad a_0+2a_3<0.
\end{equation}
In terms of the parameters used in the present work, i.e.,
\begin{equation}
\fl\qquad\quad b^+\equiv b_6,\quad b^-\equiv -b_3,\quad A_k\equiv -a_k,
\quad m^+\equiv A_0+A_2/2,\quad m^-\equiv A_0+2A_3,
\end{equation}
these restrictions become
\begin{equation}
b^+>0,\quad b^->0,\quad A_0A_2m^+>0,\quad m^->0.
\end{equation}
The Newtonian limit gives $A_0=1$. A positive kinetic term in the
action requires $A_2>0$.

In order to facilitate a comparison of the works of various groups,
we here include the parameter conversion between those of the
Cologne group of Hehl and coworkers (which we follow), Minkevich and
coworkers \cite{Min80,Min83,MN95,MG06,MGK07}, and Goenner and
M\"uller-Hoissen \cite{GFHF96}. In Hehl's work, the parameters
related to the model described in this paper are $a_0$, $a_2$,
$a_3$, $b_3$, $b_6$. Goenner and M\"uller-Hoissen used $c_1\cdots
c_9$ as the parameters in their work. The Goenner and
M\"uller-Hoissen parameters are related to Hehl's by
\begin{eqnarray}
&&c_1+3c_2=\frac{a_3}{2},\quad\sigma\equiv c_1+3c_3=\frac{a_2}{2},\quad
c_4=-\frac{a_0}{2},\nonumber\\
&&c_9-c_8=\frac{b_3}{4},\quad
c\equiv 2(6c_5+2c_6+2c_7-c_8-c_9)=\frac{b_6}{2},
\end{eqnarray}
and thus
\begin{equation}
2c_4-\sigma=m^{+},\qquad c_4-2c_1-6c_2=\frac{m^-}{2}.
\end{equation}
The parameters used by  Minkevich are related to Hehl's by
\begin{eqnarray}
b=a_3,\quad -\frac{a}{2}=a_2,\quad f_0=-\frac{a_0}{2},\quad
q_2=\frac{b_3}{4},\quad q_1=\frac{b_6-b_3}{4},\quad
f=\frac{b_6}{8},\quad
\end{eqnarray}
and thus
\begin{equation}
2f_0+\frac{a}{4}=m^{+},\quad f_0-b=\frac{m^-}{2}.
\end{equation}
\section*{References}

\end{document}